\documentclass[journal]{IEEEtran}
\IEEEoverridecommandlockouts
\usepackage{cite}
\usepackage{overpic}
\usepackage{amsmath,amssymb,amsfonts}
\usepackage{graphicx}
\usepackage{textcomp}
\usepackage{xcolor}
\usepackage{float}
\usepackage{amsthm}
\usepackage{graphicx}
\usepackage{epstopdf}
\usepackage{amsmath,bm}
\usepackage{amsfonts}
\usepackage{amssymb}
\usepackage{color}
\usepackage{multirow}
\usepackage{multicol}
\usepackage{soul,xcolor}
\usepackage{algorithm}
\usepackage{algpseudocode}%
\newcommand{\condSum}[3]{\overset{#3}{\underset{\underset{#2}{#1}}{\sum}}}

\newcommand{\mathacr}[1]{\mathsf{#1}}
\theoremstyle{plain}

\newtheorem{lemma}{Lemma}

\newtheorem{remark}{Remark}

\newcommand{\vect}[1]{\mathbf{#1}}

\def\diag{\mathrm{diag}}

\def\tr{\mathrm{tr}}

\def\Htran{\mbox{\tiny $\mathrm{H}$}}
\def\Ttran{\mbox{\tiny $\mathrm{T}$}}
\def\CN{\mathcal{N}_{\mathbb{C}}} 
\def\imagunit{\mathsf{j}} 
\def\BibTeX{{\rm B\kern-.05em{\sc i\kern-.025em b}\kern-.08em
    T\kern-.1667em\lower.7ex\hbox{E}\kern-.125emX}}
\begin{document}

\title{Is Channel Estimation Necessary to Select Phase-Shifts for RIS-Assisted Massive MIMO?
\thanks{This work was supported by the FFL18-0277 grant from the Swedish Foundation for Strategic Research. The authors are with the Department of Computer Science, KTH Royal Institute of Technology, Kista, Sweden.  (e-mail:\{ozlemtd, emilbjo\}@kth.se).}
}

\author{\IEEEauthorblockN{\"Ozlem Tu\u{g}fe Demir, \emph{Member, IEEE} and Emil Bj\"ornson, \emph{Fellow, IEEE}}}

\maketitle

\begin{abstract}
Reconfigurable intelligent surfaces (RISs) consist of many passive elements of metamaterials whose impedance can be controllable to change the characteristics of wireless signals impinging on them. Channel estimation is a critical task when it comes to the control of a large RIS when having a channel with a large number of multipath components. In this paper, we derive Bayesian channel estimators for two RIS-assisted massive  multiple-input multiple-output (MIMO) configurations: i) the short-term RIS configuration based on the instantaneous channel estimates; ii) the long-term RIS configuration based on the channel statistics. The proposed methods exploit spatial correlation characteristics at both the base station and the planar RISs, and other statistical characteristics of multi-specular fading in a mobile environment. Moreover, a novel heuristic for phase-shift selection at the RISs is developed. A computationally efficient fixed-point algorithm, which solves the max-min fairness power control optimally, is proposed. Simulation results demonstrate that the proposed uplink RIS-aided framework improves the spectral efficiency of the cell-edge mobile user equipments substantially in comparison to a conventional single-cell massive MIMO system. The impact of several channel effects are studied to gain insight about when the channel estimation, i.e., the short-term configuration, is preferable in comparison to the long-term RIS configuration to boost the spectral efficiency. 
\end{abstract}

\begin{IEEEkeywords}
RIS, massive MIMO, channel estimation, uplink spectral efficiency, max-min fair power control 
\end{IEEEkeywords}

\section{Introduction}

Reconfigurable intelligent surfaces (RISs) \cite{Huang2018a}, also known as intelligent reflecting surfaces \cite{Wu2019}, are an emerging technology for shaping the wireless medium by the software-controlled reflection of the individual propagation paths; for example, to boost the desired signal power at the receiver. The RISs are envisaged to be deployed as planar surfaces on the facades, walls, or ceilings of buildings and consist of a large number of reflecting elements \cite{RIS_Renzo_JSAC}. Each element can be made of metamaterial, acts as an isotropic scatterer when it is sub-wavelength-sized, and the impedance can be tuned to create a phase-shift pattern over the surface that reflects an incident wave as a beam in a desirable direction \cite{emil_rayleigh_fading_ris}. The aim of deploying RISs in communications is to create smart radio environments by controlling the propagation paths in both constructive and destructive ways at the desired and unintended points, respectively \cite{Bjornson2021}. RISs have been studied extensively in the literature and promising field trials have been carried out for both indoors and outdoors RIS-assisted communications \cite{RISfieldtrial}.

Since an RIS is made of passive components, the channel state information (CSI) needed to configure the RIS elements can be acquired only at the node that receives the pilot signals. To select the phase-shift of each RIS element individually in a proper way during data transmission, each channel from the transmitter to the receiver through a single element of an RIS should be estimated. However, this requires a huge number of pilot symbols in an RIS-aided wireless network. Resource-efficient channel estimation for RISs is a key open problem \cite{RIS_emil_magazine}, although many initial studies tackle the problem from different perspectives \cite{RISchannelEstimation_nested_cascaded, RISchannelEstimation_nested_LS_single,
		RISchannelEstimation_nested_iterative,
	Alwazani2020,
	RISchannelEstimation_nested_knownBSRIS2,Wei2021,Papazafeiropoulos2021,Guan2021,Chen2021,Wei2021b,Zhou2021,Wang2020,Zhaorui2020,Taha2021,Zhang2020,Kundu2021,He2021,RIS_DL_ahmet}.

Massive multiple-input multiple-output (MIMO) is the 5G technology that allows serving several user equipments (UEs) on the same time-frequency resources by spatial multiplexing \cite{massivemimobook,Marzetta2016a}. The performance of massive MIMO when it is combined with the new RIS technology has been studied in  \cite{RISmassiveMIMO_perfect_CSI,RISmassiveMIMO_LOS,RISmassiveMIMO_historical_channel,Chen2021,Zhang2020,He2021,RIS_DL_ahmet}. In the next part, we will go over each of the related works regarding massive MIMO and channel estimation in detail.

\subsection{Related Work}

In \cite{RISmassiveMIMO_perfect_CSI}, an RIS-aided massive MIMO system is considered with perfect CSI. \cite{RISmassiveMIMO_LOS} considered RIS-assisted massive MIMO with hybrid beamforming for mmWave communications. Channel estimation was not considered in that work. In \cite{Chen2021}, direction-of-angle estimation is performed to estimate the channels under the assumption that some of the RIS elements are active and have a radio frequency (RF) chain behind them. In \cite{Zhang2020}, for an RIS-assisted massive MIMO system, least squares (LS)-based channel estimation is considered with a single UE. The works \cite{He2021,RIS_DL_ahmet} utilized on/off RIS element patterns in channel estimation by exploiting sparsity. However, it is questionable if practical RIS elements can be turned off in sense of becoming full absorbers, and it leads to a large signal-to-noise ratio (SNR) loss.
In \cite{RISmassiveMIMO_historical_channel}, instead of instantaneous channel estimates, historically collected channel realizations are utilized to maximize the ergodic capacity by solving the stochastic optimization sub-problems for RIS-assisted massive MIMO.

Other than massive MIMO works, channel estimation for RIS-aided communications has also been considered in different respects. In \cite{RISchannelEstimation_nested_cascaded, RISchannelEstimation_nested_LS_single, RISchannelEstimation_nested_iterative, Alwazani2020, RISchannelEstimation_nested_knownBSRIS2} a structured phase-shift scheme is utilized in the pilot training phase to separate the estimation of different channels exploiting the orthogonality. Instead of turning off some of the RIS elements, using all of them as in those works does not lead to an SNR degradation, even such correlation always exists \cite{emil_rayleigh_fading_ris}. The works \cite{RISchannelEstimation_nested_cascaded, RISchannelEstimation_nested_LS_single} considered LS estimation without taking into account any spatial correlation among the antennas or RIS elements. In \cite{RISchannelEstimation_nested_iterative}, a tensor-based approach is adopted again without considering any spatial correlation.  The works \cite{Alwazani2020, RISchannelEstimation_nested_knownBSRIS2,Papazafeiropoulos2021} modeled the channels with spatial correlation but neglected any line-of-sight (LOS) component. In addition, they assumed a perfectly known pure LOS path for the base station (BS)-RIS channel and derived the linear minimum mean-squared error (LMMSE) channel estimator. Similarly, \cite{Kundu2021} considered spatial correlation without an LOS path. Moreover, the UE-RIS channels follow uncorrelated fading, which is physically impossible \cite{emil_rayleigh_fading_ris}. In \cite{Zhaorui2020}, the authors have also considered correlated fading for the non-deterministic BS-RIS channel and proposed a three-stage channel estimation scheme where an on/off RIS element pattern is utilized in the last stage. In this work, the LOS components of the channels are neglected. Different from these works, in \cite{Wei2021,Zhou2021} LOS is taken into account in the channel modeling but not in the LS channel estimation stage.

\subsection{Motivation and Contributions}

 In the existing RIS-assisted massive MIMO works, either perfect CSI is assumed, there are only deterministic LOS components, or previous realizations of the channels are used to optimize RIS phase-shift design. The studies that consider channel estimation in a scattering environment only considered the LS estimator, which neglects the fundamental existence of spatial correlation. In this paper, we go beyond that by developing two important components of an  RIS-assisted uplink massive MIMO system: i) channel estimation schemes for short-term (with low training overhead) and long-term RIS configurations;  ii) phase-shift design based on either small-scale channel estimates in the short-term RIS configuration or channel statistics in the long-term RIS configuration without resorting to complicated optimization. Our main contributions are:
\begin{itemize}
	\item For the first time, we consider channel estimation in an RIS-aided system by taking both the spatial correlation at the BS and multiple RISs, and random phase-shifts on the specular (dominant) paths into account, which makes the derivation different than the works \cite{Alwazani2020, RISchannelEstimation_nested_knownBSRIS2,Papazafeiropoulos2021} that assume perfectly known BS-RIS channel and neglect the LOS and any other specular components.
	\item We propose two architectures for the RIS-assisted massive MIMO in which either the short-term RIS reconfiguration based on small-scale channel estimates or the long-term RIS reconfiguration based on the channel statistics is considered.
	\item For the short-term RIS reconfiguration, we derive a Bayesian channel estimator by using a structured pilot assignment with a low training overhead. We also derive a Bayesian estimator for the channels in the case of long-term RIS reconfiguration.
	\item We propose two low-complexity closed-form phase-shift selection schemes with different overheads at the RISs and provide a respective achievable uplink spectral efficiency (SE).
\item We derive a closed-form spatial correlation matrix for a uniform planar array (UPA) using the small-angle Gaussian local scattering model. We devise a fixed-point algorithm for the optimal max-min fairness power control.
	\end{itemize}

Note that the considered pilot and phase-shift assignment for the channel estimation in case of the short-term RIS reconfiguration is similar to the structured pilot scheme used in \cite{RISchannelEstimation_nested_cascaded,RISchannelEstimation_nested_LS_single, Alwazani2020,RISchannelEstimation_nested_knownBSRIS2,RISchannelEstimation_nested_iterative}. However, these methods either ignore statistical channel knowledge, estimate only the cascaded channels, consider a single UE, assume that the BS-RIS channel is known, or consider complex iterative algorithms for channel estimation. In this paper, for the first time, we design several channel estimation schemes for the considered multi-specular spatially correlated fading (also for spatially correlated Rician fading as a specific case). Different from the conference version \cite{DemirGLOBECOM2021} that only considers channel estimation for short-term RIS configuration with only one phase-shift selection scheme, we present channel estimation for long-term RIS configuration with an additional simpler phase-shift scheme. In addition, we include the proofs of all results and the derivation of the closed-form spatial correlation matrix for a UPA.

\subsection{Paper Outline and Notations}

The rest of this paper is organized as follows. Section~\ref{sec:system-model}
    introduces the system model for the RIS-assisted massive MIMO with multi-specular correlated fading. Section~\ref{sec:phase} proposes two RIS phase-shift selection schemes that can be used with  both RIS configurations. In Section~\ref{sec:SE}, an achievable SE and receive combining schemes are presented. Section~\ref{sec:individual} reviews channel estimation with short-term RIS reconfiguration and derives a Bayesian estimator. In Section~\ref{sec:overall}, long-term RIS configuration and the respective channel estimation are considered. Section~\ref{sec:maxmin} develops a fixed-point algorithm for the optimal max-min fair power control.  Section~\ref{sec:derivation-spatial-correlation} derives a closed-form spatial correlation matrix for a UPA based on small-angle Gaussian local scattering model. In Section~\ref{sec:numerical}, the performance of the proposed frameworks are numerically evaluated and compared. Finally, the conclusions are drawn in Section~\ref{sec:conclusion}.

\emph{Notation:} Boldface lowercase and uppercase letters denote column
vectors and matrices, respectively. The superscripts $^{\Ttran}$, $^*$, $^{\Htran}$ denote transpose, conjugate, and conjugate transpose, respectively. $\vect{I}_n$ is the $n \times n$ identity matrix. The $n$-length multivariate circularly symmetric complex Gaussian distribution 
with correlation matrix $\vect{R}$ is denoted by $\CN(\vect{0}_n, \vect{R})$. The operations $\mathbb{E}\{.\}$ and $\Re\{\}$ take the expected and real values of their arguments inside, respectively.

{\bf Reproducible research:} All the simulation results can be
reproduced using the Matlab code and data files available at:
https://github.com/emilbjornson/RIS-massive-MIMO.

\section{System Model}\label{sec:system-model}

\begin{figure}[t]
	\begin{center}
		\begin{overpic}[width=7.8cm,tics=10]{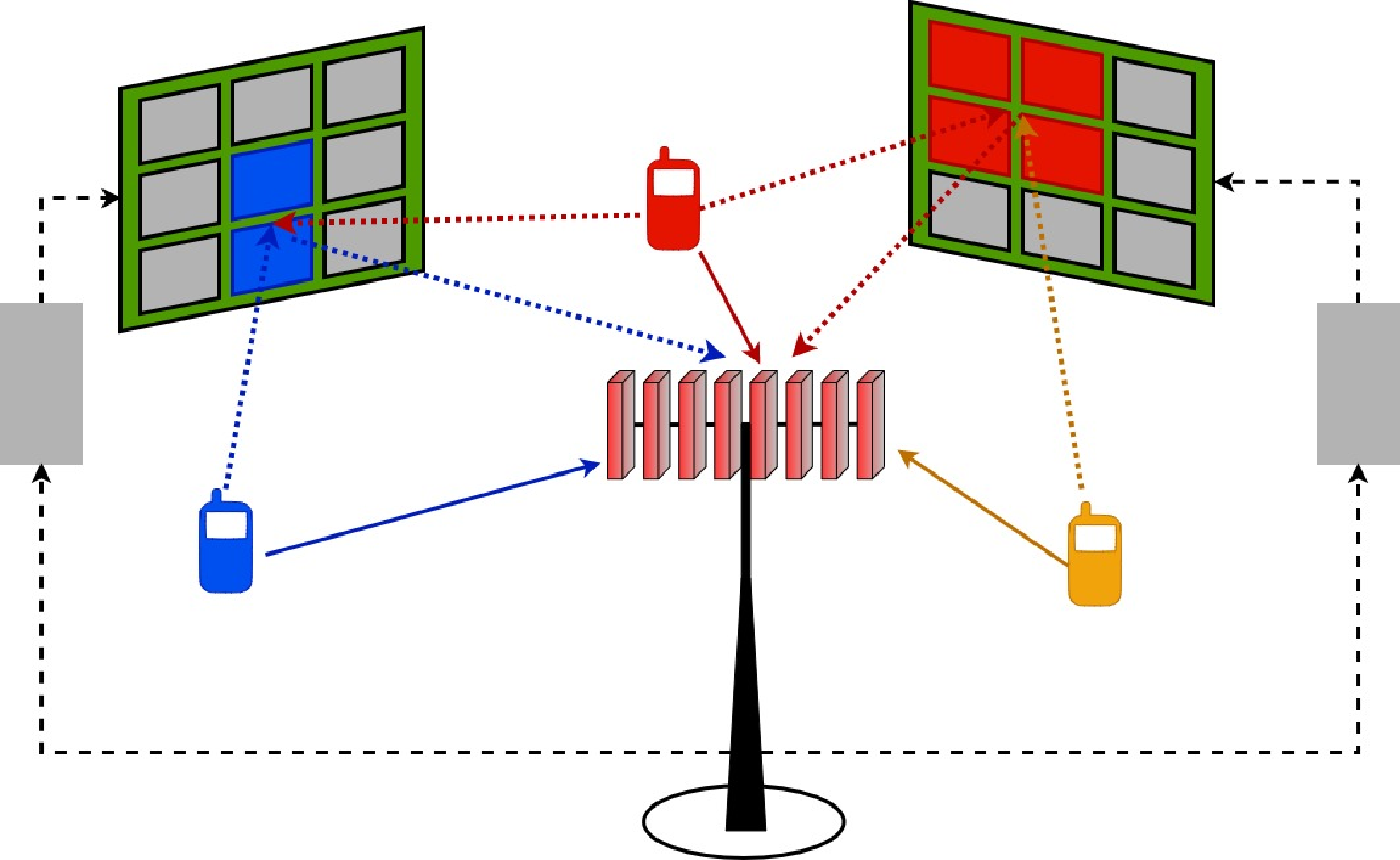}
			\put(-3.5,23){\small \rotatebox{90}{RIS Controller}}
		\put(65,4.2){\small RIS Control Link}
				\put(12,15.5){\small UE $k$}
					\put(44.5,51.5){\small UE $i$}
						\put(75,14.5){\small UE $j$}
		\put(30,20.5){$\vect{h}_k$}
		\put(51.8,40){$\vect{h}_i$}
			\put(68,21.5){$\vect{h}_j$}
		\put(33,35.2){$\vect{G}^{\prime}_k$}
		\put(62,38){$\vect{G}^{\prime}_i$}
		\put(17.5,31){$\vect{f}^{\prime}_{kk}$}
		\put(35,48){$\vect{f}^{\prime}_{ik}$}
		\put(57,51.8){$\vect{f}^{\prime}_{ii}$}
	 \put(76.5,33){$\vect{f}^{\prime}_{ji}$}
		\end{overpic}
	\end{center}
	\caption{Illustration of the considered RIS-assisted single-cell massive MIMO system model.}
	\label{fig:system} 
\end{figure}

We consider an uplink single-cell massive MIMO system that is assisted by multiple RISs, as shown in Fig.~\ref{fig:system}. The BS has $M$ antennas and each RIS has $N$ individually controllable reflecting elements. The number of RISs and UEs is $L$ and $K$, respectively. Each UE is equipped with a single antenna. To achieve low complexity in the design of the phase-shifts of RIS elements, we assume that each RIS element is assigned to one UE and its phase-shift is selected to maximize the effective channel strength of that UE. RIS assignment can be implemented based on the large-scale channel gains, and, hence the RIS elements are assigned to the UEs in the scheduling phase before data transmission. This assumption will enable us to select the RIS phase-shifts for each realization of the channels in a practical way. Let $N_k\geq 0$ denote the number of reflecting elements assigned to UE $k$, for $k=1,\ldots,K$. These elements can be located on different RISs and we have $\sum_{k=1}^KN_k=LN$ if all the RIS 
elements are assigned to the UEs. Note that $N_k$ can be zero for some UE $k$ and, in this case, no RIS element is assigned to UE $k$ to assist its communication. In between each configuration, a guard interval is needed to avoid transients. The feedback procedure can be done on a dedicated out-of-band channel that is not allocated for data transmission as discussed in \cite{Bjornson2021}.

We consider a time-varying narrowband channel. Adopting the conventional block fading model, the time resources are divided into coherence blocks \cite{massivemimobook}. The channel responses are time-invariant within each coherence block, thus represented by fixed complex scalars. We let $\tau_c$ denote the total number of samples per coherence block. Each coherence block is divided into two phases: uplink training  and uplink data transmission with $\tau_p$ and $\tau_c-\tau_p$ channel uses, respectively.

Let $\vect{h}_{k}\in \mathbb{C}^{M}$  denote the channel from UE $k$ to the BS, as shown in Fig.~\ref{fig:system}. Generalizing the spatially correlated Rician fading model \cite{DemirICASSP20} to consider an arbitrary number of specular components, each channel realization is expressed as
\begin{equation} \label{eq:channel-UE-BS}
 \vect{h}_{k}=\sum_{s=1}^{S_k^{\mathrm{h}}}e^{\imagunit\theta^{\mathrm{h}}_{k,s}}\bar{\vect{h}}_{k,s}+\tilde{\vect{h}}_k
\end{equation}
where the vector $\bar{\vect{h}}_{k,s}$ is the array steering vector scaled by the square root of the corresponding link gain for the $s$th specular component of the channel. $S_k^{\mathrm{h}}$ is the number of specular components, which are also called \emph{dominant} paths. The nonspecular part of the channel $\tilde{\vect{h}}_k$ represents the summation of other diffusely propagating multipath components. The notations used for different channels are outlined in Table~\ref{notation} for ease of readability. The vectors $\bar{\vect{h}}_{k,s}$ are fixed for a given setup. If there is an LOS path between the BS and UE $k$, then one of the specular components $e^{\imagunit\theta^{\mathrm{h}}_{k,s}}\bar{\vect{h}}_{k,s}$ corresponds to this path. Note that microscopic movements induce random phase-shifts on each individual propagation path and the one that affects the $s$th specular component of the channel is denoted by  $\theta^{\mathrm{h}}_{k,s}$. These phase-shifts change from coherence block to coherence block. Hence, they are not known in advance and modeled as independent random variables uniformly distributed in $[0,2\pi)$, i.e., $\theta^{\mathrm{h}}_{k,s} \sim \mathcal{U}[0,2\pi)$. The nonspecular paths are subject to similar phase-shifts, which give rise to small-scale fading modeled by a Gaussian distribution $\tilde{\vect{h}}_k\sim \CN(\vect{0}_M,\vect{R}^{\mathrm{h}}_k)$ that also takes an independent realization in each coherence block. The matrix  $\vect{R}^{\mathrm{h}}_k\in \mathbb{C}^{M \times M}$ describes the spatial correlation between the channel realizations observed at different BS antennas and other long-term channel effects such as geometric pathloss and shadowing. These matrices are fixed for a given setup and can, thus, be assumed to be known.

\begin{table*}[t]
\centering
	\footnotesize
\caption{Notations for the channels.}  \label{notation}
	\centering
	\begin{tabular}{|c|c|} 
		\hline
		$k$, $i$, $j$
& Indices used for UEs \\ \hline
$\ell$ & Index used for RISs \\ \hline
$n$ & Index used for RIS elements \\ \hline
$\vect{h}_{k}\in \mathbb{C}^{M}$  &    Channel from UE $k$ to the BS  \\ \hline   
	$\vect{f}_{k\ell}\in \mathbb{C}^{N}$  &   Channel from UE $k$ to the $N$ RIS elements of RIS $\ell$   \\   \hline
${\vect{f}}^{\prime}_{ki}\in \mathbb{C}^{N_i}$& Concatenated channel from UE $k$ to the $N_i$ RIS elements assigned to UE $i$ \\   \hline   
	  $\vect{G}_{\ell} \in \mathbb{C}^{M\times N}$&   Channel from RIS $\ell$ to the BS    \\   \hline  
	 $\vect{G}^{\prime}_k\in \mathbb{C}^{M\times N_k}$ &   Concatenated channel from the $N_k$ RIS elements assigned to UE $k$ to the BS    \\   \hline 
	 $\vect{H}_{k\ell}
	 =\vect{G}_{\ell}\diag(\vect{f}_{k\ell}) \in \mathbb{C}^{M \times N}$ & Channel from UE $k$ to the BS through RIS $\ell$ \\ \hline
	 $\vect{H}^{\prime}_{ij}= \vect{G}^{\prime}_{j}\diag(\vect{f}^{\prime}_{ij}) \in \mathbb{C}^{M \times N_j}$ & Concatenated channel from UE $i$ to the BS through the $N_k$ RIS elements assigned to UE $k$   \\ \hline
	\end{tabular}
\end{table*}

\begin{remark} Note that the above model is the multiple-antenna generalization and correlated fading extension of the stochastic fading model described in \cite{Jerez2019}, \cite{Chun2018}, which is also an extension of the two-way modeling with two specular components in \cite{Durgin2002}. According to the experimental results of several measurement campaigns, it has been observed that massive MIMO channels are highly directive, which leads to several dominant components in addition to the diffuse components \cite{Chun2018,nurmela2015deliverable}. When some of the channel paths cannot be modeled using Rayleigh fading due to deviating path gains, one needs to model the dominant paths separately. The channel model in  \eqref{eq:channel-UE-BS} admits correlated Rayleigh fading with $S_k^{\mathrm{h}}=0$ and Rician fading with $S_k^{\mathrm{h}}=1$ as special cases. The main motivation for this generalized model is that the dominant components of the channels (including the LOS path) mainly affect the SE of an RIS-assisted system \cite{Bjornson2021}. Unlike the previous works on massive MIMO and RIS-assisted communication works that consider only one dominant path, we consider a more general channel behavior to analyze its effect on the SE performance more deeply. 
\end{remark}

 The channel from UE $k$ to RIS $\ell$ is denoted by 
 $\vect{f}_{k\ell}\in \mathbb{C}^{N}$ and, using the same multi-specular correlated fading model as in \eqref{eq:channel-UE-BS}, it is expressed as
 \begin{equation} \label{eq:channel-UE-IRS}
  \vect{f}_{k\ell}=\sum_{s=1}^{S_{k\ell}^{\mathrm{f}}}e^{\imagunit\theta^{\mathrm{f}}_{k\ell,s}}\bar{\vect{f}}_{k\ell,s}+\tilde{\vect{f}}_{k\ell}
 \end{equation}
 where $\theta^{\mathrm{f}}_{k\ell,s} \sim \mathcal{U}[0,2\pi)$ and $\tilde{\vect{f}}_{k\ell}\sim \CN(\vect{0}_{N},\vect{R}^{\mathrm{f}}_{k\ell})$. The matrix $\vect{R}^{\mathrm{f}}_{k\ell} \in \mathbb{C}^{N\times N}$ describes the spatial correlation and pathloss corresponding to the nonspecular component of the channel. Throughout the paper, we will use the prime notation for the portions of the channels to the assigned RIS elements for a particular UE. For example, we denote the concatenated channel from UE $k$ to the $N_i$ reflecting elements that are assigned to UE $i$ out of total $LN$ RIS elements by ${\vect{f}}^{\prime}_{ki}\in \mathbb{C}^{N_i}$, as in Fig.~\ref{fig:system} and Table~\ref{notation}. This vector consists of the specific elements of $\vect{f}_{k\ell}$, $\forall  \ell$ at the corresponding indices of the assigned $N_i$ RIS elements.

 Note that the BS and the RISs are typically stationary in the envisioned use cases. Furthermore, in a well-designed system, the RISs would be deployed to have LOS paths to the BS (see \cite[Fig.~7]{Bjornson2021}). Hence, we will not consider a random phase-shift on the LOS component and express the channel between the BS and RISs by separating the LOS part. However, there can be changes in the non-line-of-sight (NLOS) paths due to the time-varying environment. Hence, we can model the channel $\vect{G}_{\ell} \in \mathbb{C}^{M\times N}$ as
\begin{equation}\label{eq:channel-BS-IRS}
\vect{G}_{\ell} = \bar{\vect{G}}_{\ell,1} + \sum_{s=2}^{S_{\ell}^{\mathrm{G}}}e^{\imagunit\theta^{\mathrm{G}}_{\ell,s}}\bar{\vect{G}}_{\ell,s}+\tilde{\vect{G}}_{\ell}
\end{equation}
which is the channel from RIS $\ell$ to the BS. Here $\bar{\vect{G}}_{\ell,1}$ is the LOS part and the second term includes the $S_{\ell}^{\mathrm{G}}-1$ specular components with random phase-shifts $\theta^{\mathrm{G}}_{\ell,s} \sim\mathcal{U}[0,2\pi)$. Using the Kronecker model \cite{Shiu2000a} with the receive and transmit correlation matrices for the BS and RISs, $\vect{R}_{\ell}^{\mathrm{G,BS}}\in \mathbb{C}^{M\times M}$ and $\vect{R}_{\ell}^{\mathrm{G,RIS}}\in \mathbb{C}^{N \times N}$, for $\ell=1,\ldots,L$, respectively, the nonspecular part of the channel $\vect{G}_{\ell}$ can be expressed as
\begin{equation}
\tilde{\vect{G}}_{\ell} = \left(\vect{R}_{\ell}^{\mathrm{G,BS}}\right)^{\frac12}\vect{W}_{\ell}\left(\vect{R}_{\ell}^{\mathrm{G,RIS}}\right)^{\frac12} \label{eq:Gtilde}
\end{equation}
where the elements of $\vect{W}_{\ell}\in \mathbb{C}^{M \times N}$ are independent and identically distributed (i.i.d.) standard complex Gaussian random variables. Let $\vect{G}^{\prime}_k\in \mathbb{C}^{M\times N_k}$ denote the concatenated channel from the subset of all $LN$ RIS elements assigned to UE $k$ to the BS, as shown in Fig.~\ref{fig:system} and Table~\ref{notation}. This matrix is constructed by picking the $N_k$ columns from the matrices $\vect{G}_{\ell}$, $\forall \ell$ corresponding to the assigned RIS elements to UE $k$.

During the uplink transmission phase, the received signal at the BS can be expressed as
\begin{align}\label{eq:received_signal} \vect{y}&=\sum_{i=1}^K\left(\vect{h}_i+\sum_{j=1}^K\vect{G}_j^{\prime}\vect{\Phi}_j\vect{f}^{\prime}_{ij}\right)s_i+\vect{n} 
\end{align}
where
$s_i$ is either the pilot or data signal of UE $i$ and $\vect{n}\sim\CN(\vect{0}_M,\sigma^2\vect{I}_M)$ is the additive noise. The phase-shift of the $n$th element of the $j$th RIS subset (assigned to UE $j$) is represented by $\phi_{j,n}\in \mathbb{C}$ with $\left\vert \phi_{j,n} \right\vert=1$, which is the $(n,n)$th element of the diagonal matrix $\vect{\Phi}_{j}\in \mathbb{C}^{N_j \times N_j}$. We consider a setup with continuous phase control, which is practically feasible \cite{RISfieldtrial}. Defining
\begin{align} \label{eq:cascaded}
\vect{H}^{\prime}_{ij}\triangleq \vect{G}^{\prime}_{j}\diag(\vect{f}^{\prime}_{ij}) \in \mathbb{C}^{M \times N_j}
\end{align}  
where $\diag(\vect{f}^{\prime}_{ij})$ denotes the $N_j\times N_j$ diagonal matrix with entries as the elements of the vector $\vect{f}^{\prime}_{ij}$, the received signal in \eqref{eq:received_signal} can also be expressed in the following form
\begin{align}\label{eq:received_signal1b} \vect{y}&=\sum_{i=1}^K\left(\vect{h}_i+\sum_{j=1}^K\vect{H}^{\prime}_{ij}\bm{\phi}_j\right)s_i+\vect{n} 
\end{align}
where $\bm{\phi}_j\in \mathbb{C}^{N_j}$ is the vector whose $n$th element is $\phi_{j,n}$. We define the composite overall channel from UE $i$ to the BS through the direct link and RISs as
\begin{align} \label{eq:bk}
\vect{b}_i&=\vect{h}_i+\sum_{j=1}^K\vect{G}_j^{\prime}\vect{\Phi}_j\vect{f}^{\prime}_{ij}=\vect{h}_i+\sum_{j=1}^K\vect{H}^{\prime}_{ij}\bm{\phi}_j, \quad i=1,\ldots,K.
\end{align}

Next, in Sections~\ref{sec:phase} and \ref{sec:SE}, we will consider different schemes for configuring the RIS phase-shifts jointly with the receive combining.  After that, we will consider two different RIS-assisted massive MIMO configurations with their respective channel estimators.

\section{Phase-Shift Selection Schemes}\label{sec:phase}

In this section, we will introduce two phase-shift selection schemes that do not require any complicated optimization algorithm and, hence, can be implemented per coherence block by using the estimates of the current channel realization. The phase-shift selection schemes will be presented using the small-scale channel estimates. However, the results in this section can be utilized for the long-term RIS configuration as we elaborate later. In the uplink, the payload data transmission phase, the received signal at the BS is given as in \eqref{eq:received_signal} or \eqref{eq:received_signal1b} with
$s_i$ being the uplink data signal of UE $i$ with transmit power $p_i$, i.e., $\mathbb{E}\{\left\vert s_i\right\vert^2\}=p_i$. 

We let $\widehat{\vect{h}}_i$ and $\widehat{\vect{H}}^{\prime}_{ij}$ denote the estimates of the direct BS-UE channel $\vect{h}_i$ and the cascaded BS-RIS-UE channel $\vect{H}^{\prime}_{ij}$ in \eqref{eq:received_signal1b}. Then, the estimate of the overall channel $\vect{b}_i$ in \eqref{eq:bk} is given as
\begin{equation} \label{eq:bk-estimate}
    \widehat{\vect{b}}_i=\widehat{\vect{h}}_i+\sum_{j=1}^K\widehat{\vect{H}}^{\prime}_{ij}\bm{\phi}_j, \quad i=1,\ldots,K.
\end{equation}

 Since the $j$th RIS subset is assigned to serve UE $j$, in the proposed schemes, $\phi_{j,n}$ will be selected to maximize the channel strength of UE $j$. More precisely, considering only the cascaded channel through the $j$th RIS subset in \eqref{eq:received_signal1b}, our aim is to maximize the norm of the respective portion of the estimated channel, i.e., $\widehat{\vect{h}}_j+\widehat{\vect{H}}^{\prime}_{jj}\bm{\phi}_j$. Under the unit modulus constraints on the elements of $\bm{\phi}_j$, the corresponding optimization problem is given as
\begin{subequations} \label{eq:optimization1}
\begin{align} 
&\underset{\bm{\phi}_j}{\textrm{maximize}} \quad \left\Vert \widehat{\vect{h}}_j+\widehat{\vect{H}}^{\prime}_{jj}\bm{\phi}_j\right\Vert^2 \label{eq:norm-maximization} \\
&\textrm{subject to} \quad \vert\phi_{j,n}\vert=1, \quad n=1,\ldots,N_j. \label{eq:norm-maximization2}\end{align}
\end{subequations}
The problem \eqref{eq:optimization1} is non-convex but semidefinite programming (SDP) with rank relaxation can be used to obtain a suboptimal solution similar to the single-UE problem in \cite[Sec.~III-A]{Wu2019}. However, SDP is computationally infeasible when having a massive number of BS antennas and RIS elements. A heuristic approximation to the optimal solution of \eqref{eq:optimization1} can be obtained by relaxing the unit modulus constraints as follows:
\begin{subequations} \label{eq:optimization2}
\begin{align} 
&\underset{\bm{\phi}_j}{\textrm{maximize}} \quad \left\Vert \widehat{\vect{h}}_j+\widehat{\vect{H}}^{\prime}_{jj}\bm{\phi}_j\right\Vert^2 \label{eq:norm-maximization3} \\
&\textrm{subject to} \quad \Vert \bm{\phi}_j\Vert^2\leq N_j. \label{eq:norm-maximization4}\end{align}
\end{subequations}
The above problem is still non-convex but it can be solved optimally by using eigenvalue decomposition and a bisection search for root finding as shown in the following lemma.
\begin{lemma}\label{lemma:norm_maximization}
	Let $\lambda_{j,d}\geq0$ be the nonnegative eigenvalues and $\vect{u}_{j,d}\in \mathbb{C}^{N_j}$ be the corresponding orthonormal eigenvectors of $(\widehat{\vect{H}}^{\prime}_{jj})^{\Htran}\widehat{\vect{H}}^{\prime}_{jj}$.
If $\widehat{\vect{h}}_j=\vect{0}_{N_j}$, the optimal solution to the problem \eqref{eq:optimization2} is given by $\bm{\phi}_j^{\star}=\sqrt{N_j}\vect{u}_{j,\bar{d}}$ where $\bar{d}$ is the index corresponding to the dominant eigenvector. Otherwise, the optimal solution is  
\begin{equation}
\bm{\phi}_j^{\star} = \sum_{d=1}^{N_j} \frac{\vect{u}_{j,d}\vect{u}_{j,d}^{\Htran}(\widehat{\vect{H}}^{\prime}_{jj})^{\Htran}\widehat{\vect{h}}_j}{\gamma^{\star}-\lambda_{j,d}}, \quad d=1,\ldots,N_j \label{eq:phi}
\end{equation}
where $\gamma^{\star}>\max_d\lambda_{j,d}$ is the unique root of
\begin{equation}
\sum_{d=1}^{N_j}\frac{\left\vert\vect{u}_{j,d}^{\Htran}(\widehat{\vect{H}}^{\prime}_{jj})^{\Htran}\widehat{\vect{h}}_j\right\vert^2}{\left(\gamma-\lambda_{j,d}\right)^2}=N_j.\label{eq:root2}
\end{equation} 
\begin{proof} The proof is given in Appendix~\ref{appendix0}.
	\end{proof}
\end{lemma}

The first scheme we propose is the one that we obtain by picking the phase-shifts of the optimal $\bm{\phi}_j^{\star}$ given in Lemma~\ref{lemma:norm_maximization}, i.e.,
\begin{equation}\label{eq:phase-shift1}
\phi_{j,n} = e^{\imagunit\angle \phi_{j,n}^{\star} }, \quad n=1,\ldots,N_j, \quad j=1,\ldots,K.
\end{equation}

\begin{remark}
For the proposed phase-shift selection scheme in \eqref{eq:phase-shift1}, we relax the original problem \eqref{eq:optimization1} and project the optimal solution of the relaxed problem \eqref{eq:optimization2} to the closest complex number on the unit circle. Although we adopt continuous phase-shift design in this paper, Lemma~\ref{lemma:norm_maximization} can also be utilized to devise a discrete phase-shift selection scheme by quantizing the optimal solution of the problem \eqref{eq:optimization2} to the closest feasible points.
\end{remark}

To obtain an even simpler scheme that avoids any eigendecomposition, we can consider 
\begin{equation} \label{eq:phase-shift3}
\phi_{j,n} = e^{\imagunit\angle{\left[\widehat{\vect{H}}^{\prime}_{jj}\right]^{\Htran}_{:n}\widehat{\vect{h}}_j}},\quad n=1,\ldots,N_j, \quad j=1,\ldots,K
\end{equation}
where $[\widehat{\vect{H}}^{\prime}_{jj}]_{:n}$ is the $n$th column of $\widehat{\vect{H}}^{\prime}_{jj}$
to maximize the norm of the estimated channel   $\widehat{\vect{h}}_j+[\widehat{\vect{H}}_{jj}^{\prime}]_{:n}\phi_{j,n}$ for each $n$ similar to the problem in \cite[Sec.~III-B]{Wu2019}. In this scheme, we approximate the norm maximization problem in \eqref{eq:optimization1} by focusing on only one cascaded path and neglecting the others.

In an aim to increase the performance, more advanced algorithms can be designed to optimize the RIS phase-shifts. However, due to more complicated algorithms, the main challenge is the computational complexity. Different from the previous works \cite{Wu2019,RISchannelEstimation_nested_knownBSRIS2} that necessitate solving a non-convex problem with computationally time-consuming algorithms, the proposed schemes in \eqref{eq:phase-shift1} and \eqref{eq:phase-shift3} do not require any heavy processing and, hence, it is computationally efficient with the same order as the linear receive combining schemes of conventional massive MIMO systems. Hence, we stick to the \emph{low-complexity processing} feature that made the canonical massive MIMO technology successful. Note that the proposed phase-shift selection schemes focus on creating strong channels and then let the BS take care of the interference suppression. The core reason is that it is easier to null interference by selecting the linear receive combiners appropriately from a convex ball of admissible vectors, than to select RIS configurations under unit-modulus constraints.  In contrast, the BS and RIS are equally capable of phase-aligning signals, which motivates our approach for the phase-shift selection.

\section{Uplink Spectral Efficiency}\label{sec:SE}

In this section, we will introduce an achievable SE expression (i.e., a lower bound on the channel capacity) for RIS-assisted massive MIMO, which is valid for any phase-shift selection, receive combining scheme, and channel estimator. The results of this section will be utilized in the simulation results to compare the SE performance of all the proposed methods. 

Let $\vect{v}_k\in \mathbb{C}^M$ denote the receive combining vector that is applied to the received signal $\vect{y}$ given in \eqref{eq:received_signal} and \eqref{eq:received_signal1b} for the decoding of the UE $k$'s uplink data at the BS. Using the composite overall channels from \eqref{eq:bk}, the soft estimate of $s_k$ that the BS obtains can be written as
\begin{equation}
\widehat{s}_k=\vect{v}_k^{\Htran}\vect{y}=\vect{v}_k^{\Htran}\vect{b}_ks_k
+\condSum{i=1}{i\neq k}{K}\vect{v}_k^{\Htran}\vect{b}_is_i+\vect{v}_k^{\Htran}\vect{n}\label{eq:soft_estimate}.
\end{equation}
A lower bound on the channel capacity, called an achievable SE, of the proposed RIS-assisted massive MIMO system can be obtained by using the classical hardening bound from the massive MIMO literature \cite[Corollary~1.3]{massivemimobook} as follows.
\begin{lemma} \label{eq:capacity1}
	An achievable SE of UE $k$ is
	\begin{equation} \label{eq:rate-expression-general}
	\mathrm{SE}_{k} = \frac{\tau_c-\tau_p}{\tau_c} \log_2  \left( 1 + \mathrm{SINR}_{k}   \right) \quad \textrm{bit/s/Hz}
	\end{equation}
	where the effective signal-to-interference-plus-noise ratio (SINR) is given by
	\begin{align} \label{eq:SINR}
	&\mathrm{SINR}_{k} =  \frac{ p_{k} \left |\mathbb{E} \left\{  \vect{v}_{k}^{\Htran} \vect{b}_k \right\}\right|^2  }{ 
		\sum\limits_{i=1}^K p_{i}  \mathbb{E} \left\{| \vect{v}_{k}^{\Htran} \vect{b}_{i} |^2\right\}
		- p_{k} \left |\mathbb{E} \left\{  \vect{v}_{k}^{\Htran} \vect{b}_{k} \right\}\right|^2  + \sigma^2 \mathbb{E} \left\{\|  \vect{v}_{k} \|^2\right\}
	}
	\end{align}
	where the expectation is with respect to the channel realizations.
\end{lemma}

\begin{proof}
	The SE in \eqref{eq:rate-expression-general} is obtained by treating  \eqref{eq:soft_estimate} as interference channel with the known channel response $\mathbb{E}\left\{\vect{v}_k^{\Htran}\vect{b}_k\right\}$ and the interference $\vect{v}_k^{\Htran}\vect{y}-\mathbb{E}\left\{\vect{v}_k^{\Htran}\vect{b}_k\right\}s_k$. Then, using \cite[Corollary~1.3]{massivemimobook} and the independence of different UE symbols and noise we obtain the given result.
\end{proof}

\begin{figure*}[t]
	\hspace{4.8cm}
		\begin{overpic}[width=9.2cm,tics=10]{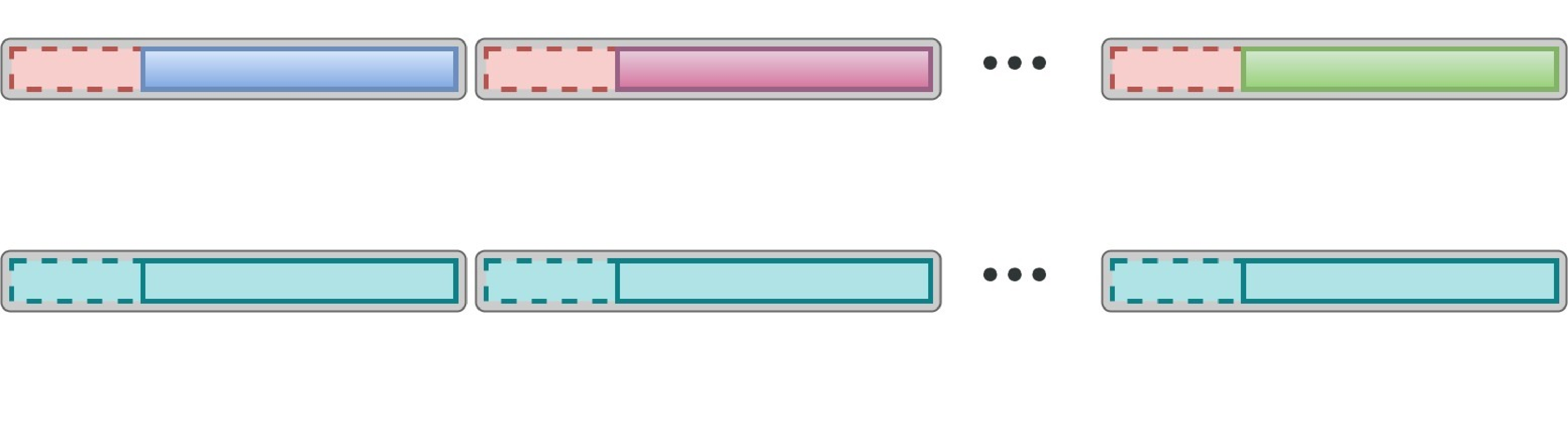}
				\put(-44,23.5){\footnotesize Short-term RIS configuration}
			
	\put(-44,9.5){\footnotesize Long-term RIS configuration}
	
			\put(1.5,19){\footnotesize Pilot}
			\put(15.5,19){\footnotesize Data}
				\put(32,19){\footnotesize Pilot}
			\put(46,19){\footnotesize Data}
				\put(72,19){\footnotesize Pilot}
			\put(86,19){\footnotesize Data}
					\put(1.5,5.5){\footnotesize Pilot}
			\put(15.5,5.5){\footnotesize Data}
				\put(32,5.5){\footnotesize Pilot}
			\put(46,5.5){\footnotesize Data}
				\put(72,5.5){\footnotesize Pilot}
			\put(86,5.5){\footnotesize Data}
	
		\end{overpic}
	\caption{Two different RIS configurations for a given long-term channel statistics. Each main block represents a coherence block. The same RIS phase-shift configuration is used when the color of the two sub-blocks are the same.}
	\label{fig:channel_estimation_schemes}
\end{figure*} 

Note that Lemma~\ref{eq:capacity1} is valid for any selection of receive combining vector $\vect{v}_k$, based on any type of channel estimates that are independent of the data. Maximum ratio (MR) combining can be used to maximize the received power, which for the considered RIS-assisted massive MIMO system can be defined as
\begin{equation}\label{eq:MR}
\vect{v}_k^{\mathrm{MR}}= \widehat{\vect{b}}_k
\end{equation}
where we recall that  $\widehat{\vect{b}}_k$ is the estimate of the overall channel $\vect{b}_k$ from \eqref{eq:bk-estimate}, obtained using one of the schemes that will be described in later sections. Since MR ignores the interference, and there can exist substantial interference in the considered setup, we will also consider the regularized zero forcing (RZF) scheme that actively suppresses interference using the combining vector
\begin{equation}\label{eq:RZF}
\vect{v}_k^{\mathrm{RZF}}=\left(\sum_{i=1}^Kp_i\widehat{\vect{b}}_i\widehat{\vect{b}}_i^{\Htran}+\sigma^2\vect{I}_M\right)^{-1}\widehat{\vect{b}}_k.
\end{equation}

In the following sections, we will propose different RIS configurations leading to different channel estimation and RIS phase-shift selection methods, and uplink SEs. The considered two different RIS configurations are depicted in Fig.~\ref{fig:channel_estimation_schemes} for given long-term channel statistics that are the same throughout many coherence blocks. When the same color is used for different sub-blocks in this figure, it is understood that the RISs adopt the same phase-shift configuration. For the short-term RIS configuration, in the pilot transmission phase, a predefined orthogonal phase-shift structure is adopted for the RISs whose details will be provided in the next section. Based on received pilot signals, the individual BS-RIS-UE and direct BS-UE channels are estimated and then used in the RIS phase-shift design. RIS phase-shifts are reconfigured in the data transmission phase based on small-scale fading. For the second scheme, the channel statistics are utilized to configure the RIS phase-shifts, and, hence throughout the communication, the same configuration is adopted. Using the received pilot signals, the overall UE-BS channels are estimated. For this scheme, the estimated channels are not used for RIS configuration. Note that Fig.~2 is just for visualization purpose and the length of pilot and data blocks can differ for both RIS configurations.

\section{Channel Estimation for The Short-Term RIS Configuration} \label{sec:individual}

The RIS phase-shifts in \eqref{eq:phase-shift1} or \eqref{eq:phase-shift3} depend on estimates of the direct channels $\vect{h}_i$ and the cascaded RIS channels $\vect{H}^{\prime}_{ij}$. In this section, we will describe a way to estimate these channels in each coherence block, so that the phase-shifts and receive combiner can be changed in every block, based on the small-scale fading variations. The Bayesian channel estimator derived in this section is utilized in the training stage of the short-term RIS configuration in Fig.~\ref{fig:channel_estimation_schemes}. For the proposed channel estimation scheme, the number of samples that are allocated for pilot transmission in the training phase is  $\tau_p=(LR+1)K$ under the assumption that $\tau_p<\tau_c$. Here, $1\leq R\leq N$ is a predefined integer parameter such that $N/R$ is also an integer. $R$ is equal to the number of non-overlapping RIS sub-surfaces with $N/R$ elements in a given RIS, which are introducing the same phase-shifts in the pilot training phase. For the proposed method, the parameter $R$ can be adjusted arbitrarily. The value of $R$ for a given setup can be adjusted by considering the trade-off between the pilot length and the channel estimation performance. Note that a fixed deterministic RIS phase-shift pattern is utilized during pilot transmission as shown in Fig.~2 and the BS has the knowledge of this pattern to despread the received pilots signals.

\begin{figure*}[t]
\centering
		\begin{overpic}[width=9.6cm,tics=10]{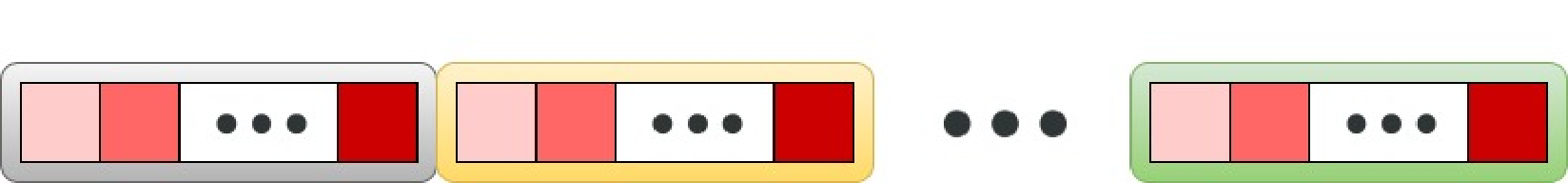}
			\put(-19,10){\small RIS $(\ell,r)$}
			\put(-19,-14){\small Received}
			\put(-21,-18.5){\small pilot signals}
			\put(9.5,10){ $\psi_{\ell r,0}$}
			\put(37,10){ $\psi_{\ell r,1}$}
			\put(79.5,10){ $\psi_{\ell r,LR}$}
			\put(-16,-5){\small UE $k$}
			\put(10.5,-4.5){ $\bm{\varphi}_k$}
			\put(12,-16){\small $\vect{Y}^{\mathrm{p}}_{0}$}
			\put(40,-16){\small $\vect{Y}^{\mathrm{p}}_{1}$}
			\put(83.5,-16){\small $\vect{Y}^{\mathrm{p}}_{LR}$}			
			\put(38.5,-4.5){ $\bm{\varphi}_k$}
			\put(83,-4.5){ $\bm{\varphi}_k$}
			\put(5,-9){\small$\varphi_{k,1} \ldots\varphi_{k,K}$}
			\put(33,-9){\small$\varphi_{k,1} \ldots\varphi_{k,K}$}
			\put(77.5,-9){\small$\varphi_{k,1} \ldots\varphi_{k,K}$}
		\end{overpic}
	\vspace{18mm}
	\caption{Pilot structure and RIS phase-shifts in Section~\ref{sec:individual}.}
	\label{fig:channel_estimation}
\end{figure*}

We consider $LR+1$ different pilot transmission intervals, each spanning $K$ channel uses, as depicted in Fig.~\ref{fig:channel_estimation}. Let $\bm{\varphi}_k\in \mathbb{C}^{K}$ denote the pilot signal assigned to UE $k$ for each training interval where $\Vert\bm{\varphi}_{k}\Vert^2=1$, $\forall k$. The pilot signals are mutually orthogonal between UEs, i.e., $\bm{\varphi}_k^{\Htran}\bm{\varphi}_i=0$, $\forall i\neq k$. We will use the index $t=0$ for the first training interval and $t=1,\ldots,LR$ for the consecutive $LR$ intervals. The phase-shift introduced by each element in the $r$th sub-surface of RIS $\ell$ in the $t$th pilot interval is represented by $\psi_{\ell r,t}\in\mathbb{C}$ with $\vert \psi_{\ell r,t}\vert=1$, for $\ell=1,\ldots,L$, $r=1,\ldots,R$, and $t=0,\ldots,LR$. Let RIS $(\ell,r)$ refer to the elements in the $r$th sub-surface of RIS $\ell$ as in Fig.~\ref{fig:channel_estimation}. Let also $\bm{\psi}_{\ell r}\in \mathbb{C}^{LR+1}$ denote the vector that is constructed by the phase-shifts of the RIS $(\ell,r)$ throughout the whole training interval, i.e., $\bm{\psi}_{\ell r}=[\psi_{\ell r,0} \ \ldots \ \psi_{\ell r,LR}]^{\Ttran}\in \mathbb{C}^{LR+1}$, for $\ell=1,\ldots,L$ and $r=1,\ldots,R$. We select them so that they are mutually orthogonal and also orthogonal to the all-ones vector ${\bf 1}_{LR+1}=[ 1\ \ldots\ 1]^{\Ttran}\in \mathbb{C}^{LR+1}$, i.e.,  $\bm{\psi}_{\ell r}^{\Htran}\bm{\psi}_{\ell^{\prime}r^{\prime}}=0$, $\forall (\ell^{\prime},r^{\prime}) \neq (\ell,r)$ and $\bm{\psi}_{\ell r}^{\Htran}{\bf 1}_{LR+1}=0$, for  $\forall \ell, \forall r$.  This is satisfied by the $LR$ columns of the $(LR+1)\times (LR+1)$ discrete Fourier transform (DFT) matrix by excluding the first column (which is the all-ones vector). This particular selection will enable to separate the received pilot signals and thereby simplify the channel estimation. To make the notation simpler, let us first define
\begin{equation} \label{eq:cascaded2}
\vect{H}_{k\ell}\triangleq \vect{G}_{\ell}\diag(\vect{f}_{k\ell}) \in \mathbb{C}^{M \times N}
\end{equation}  
in analogy with the definition in \eqref{eq:cascaded}. The columns of the matrix $\vect{H}_{k\ell}$ correspond to the cascaded channels from UE $k$ to the BS through RIS $\ell$. Let also $\mathrm{RIS}_{\ell,r}$ denote the set of element indices corresponding to the RIS $(\ell,r)$. The channels $[\vect{H}_{k\ell}]_{:n}$ for $n\in \mathrm{RIS}_{\ell,r}$ will be affected by the same phase-shift $\psi_{\ell r,t}$ in the $t$th training interval. Using these definitions, the received signal at the BS in the $t$th interval is given as
\begin{align}
\vect{Y}^{\mathrm{p}}_{t}=&\sum_{i=1}^K\sqrt{K\eta}\left(\vect{h}_i+\sum_{\ell=1}^{L}\sum_{r=1}^{R} \psi_{\ell r,t}\sum_{n\in \mathrm{RIS}_{\ell,r}}\left[\vect{H}_{i\ell}\right]_{:n}\right)\bm{\varphi}_{i}^{\Ttran}\nonumber\\
&+\vect{N}^{\mathrm{p}}_{t} \label{eq:received_pilot}
\end{align}
where $\vect{N}_t^{\mathrm{p}}\in \mathbb{C}^{M\times K}$ is the additive noise with i.i.d. $\CN(0,\sigma^2)$ elements. The pilot transmit power is denoted by $\eta$. 
After correlating   $\vect{Y}^{\mathrm{p}}_t$ with $\bm{\varphi}_{k}$, we obtain the following sufficient statistics for the estimation of $\vect{h}_k$ and $\vect{H}_{k\ell}$, for $\ell=1,\ldots,L$:
\begin{align}
\vect{z}_{k,t}^{\mathrm{p}}&=\vect{Y}^{\mathrm{p}}_t\bm{\varphi}_{k}^*\nonumber\\
&=\sqrt{K\eta}\left(\vect{h}_k+\sum_{\ell=1}^{L}\sum_{r=1}^{R}\psi_{\ell r,t}\sum_{n\in \mathrm{RIS}_{\ell,r}}\left[\vect{H}_{k\ell}\right]_{:n}\right)+\tilde{\vect{n}}_{k,t}^{\mathrm{p}} \label{eq:sufficient_statistics_k}
\end{align}
where $\tilde{\vect{n}}_{k,t}^{\mathrm{p}}=\vect{N}_t^{\mathrm{p}}\bm{\varphi}_{k}^*\sim\CN(\vect{0}_M,\sigma^2\vect{I}_{M})$. Since ${\bf 1}_{LR+1}$ is orthogonal to the vectors $\bm{\psi}_{\ell r}$ by construction, we have ${\bf 1}_{LR+1}^{\Htran}\bm{\psi}_{\ell r}= \sum_{t=0}^{LR}\psi_{\ell r,t}=0$, for $\ell=1,\ldots,L$ and $r=1,\ldots,R$. We can obtain the sufficient statistics for the estimation of direct channel $\vect{h}_k$ using all the received pilot signals as follows:
\begin{equation}
\vect{z}_{k}^{\mathrm{p,h}}=\frac{\sum_{t=0}^{LR}\vect{z}_{k,t}^{\mathrm{p}}}{\sqrt{LR+1}}=\sqrt{K(LR+1)\eta}\vect{h}_k+\tilde{\vect{n}}_{k}^{\mathrm{p,h}} \label{eq:sufficient_statistics_k_0}
\end{equation}
where $\tilde{\vect{n}}_{k}^{\mathrm{p,h}}=\sum_{t=0}^{LR}\tilde{\vect{n}}_{k,t}^{\mathrm{p}}\big/\sqrt{LR+1} \sim\CN(\vect{0}_M,\sigma^2\vect{I}_{M})$. Note that the pilots and phase-shifts are designed to enable separate estimation of the channels.

The LMMSE estimate of $\vect{h}_k$ is given by \cite{Kay1993a}
\begin{equation}\label{eq:hk_hat}
\widehat{\vect{h}}_{k} \triangleq \sqrt{K(LR+1)\eta}\overline{\vect{R}}_k^{\mathrm{h}}\left(K(LR+1)\eta\overline{\vect{R}}_k^{\mathrm{h}}+\sigma^2\vect{I}_M\right)^{-1}\vect{z}_{k}^{\mathrm{p,h}}
\end{equation}
where it follows from \eqref{eq:channel-UE-BS} that $\overline{\vect{R}}_k^{\mathrm{h}}$ can be computed as
\begin{equation}
\overline{\vect{R}}_k^{\mathrm{h}}=\mathbb{E}\left\{\vect{h}_k\vect{h}_k^{\Htran}\right\}=\sum_{s=1}^{S_k^{\mathrm{h}}}\bar{\vect{h}}_{k,s}\bar{\vect{h}}_{k,s}^{\Htran}+\vect{R}_k^{\mathrm{h}}. \label{eq:Rhbar}
\end{equation}

The sufficient statistics for the estimation of $[\vect{H}_{k\ell}]_{:n}$, for $n\in \mathrm{RIS}_{\ell,r}$ is obtained by combining the signals \eqref{eq:sufficient_statistics_k} with appropriate weighting as
\begin{align}
\vect{z}_{k,\ell r}^{\mathrm{p,H}}&=\frac{\sum_{t=0}^{LR}\psi_{\ell r,t}^{*}\vect{z}_{k,t}^{\mathrm{p}}}{\sqrt{LR+1}}\nonumber\\&=\sqrt{K(LR+1)\eta}\sum_{n\in \mathrm{RIS}_{\ell,r}}\left[\vect{H}_{k\ell}\right]_{:n}+\tilde{\vect{n}}_{k,\ell r}^{\mathrm{p,H}} \label{eq:sufficient_statistics_k_ell}
\end{align}
where we have used that $\bm{\psi}_{\ell^{\prime} r^{\prime}}^{\Htran}\bm{\psi}_{\ell r}=0$, $\forall (\ell^{\prime},r^{\prime})\neq (\ell,r)$ and  $\bm{\psi}_{\ell r}^{\Htran}{\bf 1}_{LR+1}=0$,  $\forall \ell, \forall r$, and $\tilde{\vect{n}}_{k,\ell r}^{\mathrm{p,H}}=\sum_{t=0}^{LR}\psi_{\ell r,t}^*\tilde{\vect{n}}_{k,t}^{\mathrm{p}}\big/\sqrt{LR+1} \sim\CN(\vect{0}_M,\sigma^2\vect{I}_{M})$. 
\begin{lemma}\label{lemma1}
	The LMMSE estimate of $\left[\vect{H}_{k\ell}\right]_{:n}$, for $n\in \mathrm{RIS}_{\ell,r}$ is given by
	\begin{align}
	\left[\widehat{\vect{H}}_{k\ell}\right]_{:n} \triangleq& \sqrt{K(LR+1)\eta}\overline{\vect{R}}_{k\ell,n} \nonumber\\
	&\times \left(K(LR+1)\eta\overline{\vect{R}}_{k(\ell,r)}
	+\sigma^2\vect{I}_{M}\right)^{-1}\vect{z}_{k,\ell r}^{\mathrm{p,H}} \label{eq:f_kell_estimates}
	\end{align}
	where 
	\begin{align}
	\overline{\vect{R}}_{k\ell,n}=&\sum_{n^{\prime}\in \mathrm{RIS}_{\ell,r}}\left[\overline{\vect{R}}_{k\ell}^{\mathrm{f}}\right]_{nn^{\prime}}\Bigg(\sum_{s=1}^{S_{\ell}^{\mathrm{G}}} \left[\bar{\vect{G}}_{\ell,s}\right]_{:n}\left[\bar{\vect{G}}_{\ell,s}\right]_{:n^{\prime}}^{\Htran}\nonumber\\
	&+\left[\vect{R}_{\ell}^{\mathrm{G,RIS}}\right]_{n^{\prime}n}\vect{R}_{\ell}^{\mathrm{G,BS}} \Bigg), \\
	\overline{\vect{R}}_{k(\ell,r)} =&\sum_{n\in \mathrm{RIS}_{\ell,r}}\overline{\vect{R}}_{k\ell,n}
	\end{align}
	with
	\begin{equation}
	\overline{\vect{R}}_{k\ell}^{\mathrm{f}}=\mathbb{E}\left\{\vect{f}_{k\ell}\vect{f}_{k\ell}^{\Htran}\right\}=\sum_{s=1}^{S_{k\ell}^{\mathrm{f}}}\bar{\vect{f}}_{k\ell,s}\bar{\vect{f}}_{k\ell,s}^{\Htran}+\vect{R}_{k\ell}^{\mathrm{f}}. \label{eq:Rfbar2}
	\end{equation} 
	\begin{proof}
		The proof is given in Appendix~\ref{appendix1}.
	\end{proof}
\end{lemma}

As can be seen from Lemma~\ref{lemma1}, we exploit not only the LOS part of the BS-RIS channels $\vect{G}_{\ell}$ but also the other dominant components and the spatial correlation characteristics at both the RIS and BS sides. This makes the proposed channel estimation scheme novel and more advanced than the existing schemes (see the introduction for details). 

Note that the RIS subsets of elements share the same phase-shift during only the pilot transmission phase. Utilizing the spatial correlation structure of a planar RIS, Lemma~3 demonstrates how each individual channel can be estimated. Later, in data transmission phase, each RIS can have potentially different phase-shifts using the proposed phase-shift selection methods.

Using the estimated channels, we adjust the RIS phase-shifts for the data transmission according to \eqref{eq:phase-shift1} or \eqref{eq:phase-shift3}. Using those phase-shifts, the estimate of the overall channel $\vect{b}_k$ in \eqref{eq:bk} is obtained as
\begin{equation} \label{eq:bhat}
\widehat{\vect{b}}_k=\widehat{\vect{h}}_k+\sum_{j=1}^K\widehat{\vect{H}}^{\prime}_{kj}\bm{\phi}_j, \quad k=1,\ldots,K
\end{equation}
where the cascaded channel estimates $\widehat{\vect{H}}^{\prime}_{kj}$ can be obtained by picking $[\widehat{\vect{H}}_{k\ell}]_{:n}$ in \eqref{eq:f_kell_estimates} as its columns according to the definition and indexing in \eqref{eq:cascaded}. Once we have obtained $\widehat{\vect{b}}_k$ from \eqref{eq:bhat}, the receive combining schemes given in \eqref{eq:MR} and \eqref{eq:RZF} can be used with the achievable SE expression presented in Lemma~\ref{eq:capacity1}.

\section{Channel Estimation For The Long-Term Phase-Shift Configuration} \label{sec:overall}

The previous section described how to estimate the individual channels in each coherence block to enable reconfiguration of the RIS phase-shifts and combining vectors in every coherence block (short-term configuration). In this section, we will explore another option: select the RIS phase-shifts based on long-term statistics and keep the configuration fixed for several consecutive coherence blocks. In this case, only the effective channel obtained with those phase-shifts need to be estimated and used for receive combining. The Bayesian estimator derived in this section is utilized in the long-term RIS configuration (see Fig.~\ref{fig:channel_estimation_schemes}). To this end, we will propose a heuristic phase-shift selection scheme, which make use of the correlation matrices, and describe the remaining channel estimation problem.

\subsection{Phase-Shift Selection Based on the Channel Correlation Matrices}

In a practical setup, it is straightforward for the BS to acquire the correlation matrices $\overline{\vect{R}}_k^{\mathrm{h}}$ in \eqref{eq:Rhbar} and $\overline{\vect{R}}_{k\ell}^{\mathrm{f}}$ in \eqref{eq:Rfbar2} using  sample covariance matrix estimation techniques. Assuming that these matrices are available, we can select the phase-shifts using the methods in Section~\ref{sec:phase}  based on the dominant eigenvectors of the above correlation matrices, i.e.,
\begin{equation} \label{eq:phase_shifts3}
\widehat{\vect{H}}^{\prime}_{jj} = \bar{\vect{G}}^{\prime}_{j,1}\diag\left(\breve{\vect{f}}_{jj}^{\prime}\right), \quad \widehat{\vect{h}}_j=\breve{\vect{h}}_{j}
\end{equation}
by which we replace the small-scale channel estimates in Section~III. In \eqref{eq:phase_shifts3}, $\breve{\vect{f}}_{jj}^{\prime}\in \mathbb{C}^{N_j}$ is the portion of the dominant eigenvectors of $\overline{\vect{R}}_{j\ell}^{\mathrm{f}}$ corresponding to the RIS elements that are assigned to UE $j$. The vector  $\breve{\vect{h}}_{j}\in \mathbb{C}^M$ is the dominant eigenvector of $\overline{\vect{R}}_j^{\mathrm{h}}$. Such a phase-shift selection can be fixed for a long time. Note that the dominant eigenvectors of the correlation matrices will be equal to LOS paths when they exist and dominate the other channel components. 

\subsection{Estimation of the Cascaded Channel}

Now, assuming a fixed phase-shift configuration, that can be selected according to  \eqref{eq:phase_shifts3}, we will consider estimation of the resulting effective channel between the UEs and the BS. 
We begin by re-expressing the received signal in \eqref{eq:received_signal} as follows:
\begin{equation}\label{eq:received_signal2} \vect{y}=\sum_{i=1}^K \underbrace{\left(\vect{h}_i+\sum_{\ell=1}^{L}\vect{G}_{\ell}\overline{\vect{\Phi}}_{\ell}\vect{f}_{i\ell}\right)}_{ \vect{b}_i} s_i+\vect{n}
\end{equation}
where $\overline{\vect{\Phi}}_{\ell}\in \mathbb{C}^{N \times N}$ is the diagonal matrix with the $(n,n)$th diagonal element being the phase-shift of the $n$th element of RIS $\ell$. 
Since the phase-shifts are constant, the effective channel $\vect{b}_i \in \mathbb{C}^M$ in \eqref{eq:received_signal2} is not controllable anymore and can be estimated as single vector, without separating it into its individual components.

To estimate $\vect{b}_1,\ldots,\vect{b}_K$, suppose the uplink training length is $\tau_p=\dot{\tau}K$ channel uses and each UE is assigned a unique pilot sequence from a set of $K$ mutually orthogonal pilot sequences of length $\tau_p$. Here, $\dot{\tau}\geq1$ is an arbitrary integer and, in fact, $\dot{\tau}=1$ is enough to guarantee orthogonality between different UE pilot sequences. However, we will keep it as a design parameter to achieve a fair comparison between different schemes considered in this paper. 

Similar to the previous channel estimation scheme, let $\bm{\dot{\varphi}}_k\in \mathbb{C}^{\dot{\tau}K}$ denote the pilot signal assigned to UE $k$ where $\Vert\bm{\dot{\varphi}}_{k}\Vert^2=1$, $\forall k$. The pilot signals are mutually orthogonal, i.e., $\bm{\dot{\varphi}}_k^{\Htran}\bm{\dot{\varphi}}_i=0$, $\forall i\neq k$. The received signal at the BS in the training interval is given as
\begin{equation}\label{eq:received_pilot2}
\vect{Y}^{\mathrm{p}}=\sum_{i=1}^K\sqrt{\dot{\tau}K\eta} \vect{b}_i \bm{\dot{\varphi}}_{i}^{\Ttran}+\vect{N}^{\mathrm{p}}
\end{equation}
where $\vect{N}^{\mathrm{p}}\in \mathbb{C}^{M\times \dot{\tau}K}$ is the additive noise with i.i.d. $\CN(0,\sigma^2)$ elements. The pilot transmit power is denoted by $\eta$ as in \eqref{eq:received_pilot}. After correlating $\vect{Y}^{\mathrm{p}}$ with $\bm{\dot{\varphi}}_{k}$, we obtain the sufficient statistics for the estimation of the overall channel $\vect{b}_k$ as
\begin{equation}
\vect{z}_{k}^{\mathrm{p}}=\vect{Y}^{\mathrm{p}}\bm{\dot{\varphi}}_{k}^*=\sqrt{\dot{\tau}K\eta}\vect{b}_k+\tilde{\vect{n}}_{k}^{\mathrm{p}} \label{eq:sufficient_statistics_k2}
\end{equation}
where $\tilde{\vect{n}}_{k}^{\mathrm{p}}=\vect{N}^{\mathrm{p}}\bm{\dot{\varphi}}_{k}^*\sim\CN(\vect{0}_M,\sigma^2\vect{I}_{M})$.  The LMMSE estimate of $\vect{b}_{k}$ is obtained as follows.
\begin{lemma}\label{lemma3}
	The LMMSE estimate of $\vect{b}_{k}$ is given by
	\begin{equation}
	\widehat{\vect{b}}_{k} \triangleq \sqrt{\dot{\tau}K\eta}\overline{\vect{R}}_{k}^{\rm{b}}  \left(\dot{\tau}K\eta\overline{\vect{R}}_{k}^{\rm{b}}  +\sigma^2\vect{I}_{M}\right)^{-1}\vect{z}_{k}^{\mathrm{p}} \label{eq:h_kdot_estimates}
	\end{equation}
	where $\overline{\vect{R}}_{k}^{\rm{b}} =\mathbb{E}\{\vect{b}_k\vect{b}_k^{\Htran}\}$ is given as
\begin{align}
\overline{\vect{R}}_{k}^{\rm{b}}  =& \overline{\vect{R}}_k^{\mathrm{h}}+ \sum_{\ell=1}^L \sum_{s=1}^{S_{\ell}^{\mathrm{G}}} \bar{\vect{G}}_{\ell,s}\overline{\vect{\Phi}}_{\ell}\overline{\vect{R}}_{k\ell}^{\mathrm{f}}\overline{\vect{\Phi}}_{\ell}^{\Htran}\bar{\vect{G}}_{\ell,s}^{\Htran}\nonumber\\
&+\sum_{\ell=1}^L\tr\left(\vect{R}_{\ell}^{\mathrm{G,RIS}}\overline{\vect{\Phi}}_{\ell}\overline{\vect{R}}_{k\ell}^{\mathrm{f}}\overline{\vect{\Phi}}_{\ell}^{\Htran}\right)\vect{R}_{\ell}^{\mathrm{G,BS}}.
\end{align}
	\begin{proof}
		The proof follows from the standard estimation theory, computing $\overline{\vect{R}}_{k}^{\rm{b}} =\mathbb{E}\{\vect{b}_k\vect{b}_k^{\Htran}\}$ using the statistical properties of the individual channels, and following similar steps as in \eqref{eq:expectation}.
	\end{proof}
\end{lemma}

The channel estimates $	\widehat{\vect{b}}_{k}$ in \eqref{eq:h_kdot_estimates} can now be utilized to select the receive combining vectors $\vect{v}_k$.  The corresponding  SE can be computing using Lemma~\ref{eq:capacity1}. The benefit of the scheme described in this section is that the required number of pilot symbols to guarantee orthogonality can be greatly reduced, while the drawback is that selecting the phase-shifts based on only long-term statistics might lead to an SE reduction.

\section{Max-Min Fairness Power Control}\label{sec:maxmin}

In an RIS-assisted massive MIMO system, the most unfortunate UEs with blocked direct channels to the BS or severe pathlosses will be more likely to gain from the phase-shifts introduced by RISs to maximize their SNRs. To quantify the performance improvement over conventional massive MIMO, the smallest SEs in the network are thus of great importance. We will therefore focus on the max-min fairness metric.
To achieve a fair comparison between the two systems, it is important they are both optimized towards the considered metric.

In this section, we will propose a fixed-point algorithm for the max-min fairness-based power control, where the aim is to maximize the minimum SE among all the UEs. The variables of the considered optimization problem are the uplink transmit powers of UEs, i.e., $p_k$, for $k=1,\ldots,K$. Noting that maximizing the minimum of the achievable SEs in \eqref{eq:rate-expression-general} is equivalent to maximizing the minimum of the SINRs in \eqref{eq:SINR}, by the monotonicity of the logarithm, the max-min fairness power control problem can be cast as

\begin{subequations} \label{eq:maxminfairopt}
\begin{align}
& \underset{\left\{p_k:k=1,\ldots,K\right\}}{\mathacr{maximize}} \ \ \underset{k\in\{1,\ldots,K\}}{\mathacr{min}}  \nonumber\\
&\frac{ p_{k} \left |\mathbb{E} \left\{  \vect{v}_{k}^{\Htran} \vect{b}_k \right\}\right|^2  }{ 
	\sum\limits_{i=1}^K p_{i}  \mathbb{E} \left\{| \vect{v}_{k}^{\Htran} \vect{b}_{i} |^2\right\}
	- p_{k} \left |\mathbb{E} \left\{  \vect{v}_{k}^{\Htran} \vect{b}_{k} \right\}\right|^2  + \sigma^2 \mathbb{E} \left\{\|  \vect{v}_{k} \|^2\right\}
}   \label{eq:objective}\\
& {\mathacr{subject \ to}} \quad  0\leq p_k\leq p_{\rm max}, \ \   k=1, \ldots, K \label{eq:constraint-maxmin}
\end{align}	
\end{subequations}
where $p_{\rm max}$ is the maximum uplink data transmission power for each UE. Note that the above optimization problem can be solved optimally via bisection search. At each iteration of the bisection search, a linear feasibility problem needs to be solved similar to the classical massive MIMO power allocation problems \cite[Sec.~7.1]{massivemimobook}. Although linear feasibility problems are relatively easy to handle compared to more complicated non-linear optimization problems, bisection search is still computationally demanding since multiple of feasibility checks are implemented. An alternative approach is to exploit the specific structure of the problem for a computationally efficient fixed-point algorithm. We will use the following lemma to solve the above problem optimally using a simple fixed-point algorithm.

\begin{lemma} \label{lemma:convergence}
	The fixed-point algorithm whose steps are outlined in Algorithm~\ref{alg:fixed-point} converges to the optimal solution of \eqref{eq:maxminfairopt}.
\end{lemma}
\begin{proof}
	The proof follows from \cite[Lemma~1, Theorem~1]{Hong2014}.
\end{proof} 

\begin{algorithm}
	\caption{Fixed-point algorithm for solving the max-min fairness problem in \eqref{eq:maxminfairopt}.} \label{alg:fixed-point}
	\begin{algorithmic}[1]
		\State {\bf Initialization:} Set arbitrary $p_k>0$, for $k=1,\ldots,K$, and the solution accuracy $\varepsilon>0$.
		\While{$\underset{k\in\{1,\ldots,K\}}{\max} \textrm{SINR}_k-\underset{k\in\{1,\ldots,K\}}{\min} \textrm{SINR}_k > \varepsilon$} 
		\State $p_k \gets \frac{ \sum\limits_{i=1}^K p_{i}  \mathbb{E} \left\{| \vect{v}_{k}^{\Htran} \vect{b}_{i} |^2\right\}
			- p_{k} \left |\mathbb{E} \left\{  \vect{v}_{k}^{\Htran} \vect{b}_{k} \right\}\right|^2  + \sigma^2 \mathbb{E} \left\{\|  \vect{v}_{k} \|^2\right\}}{\left |\mathbb{E} \left\{  \vect{v}_{k}^{\Htran} \vect{b}_k \right\}\right|^2  }$, $\quad$ $k=1,\ldots,K.$
		\State $\widetilde{p} \gets \underset{k\in\{1,\ldots,K\}}{\max}p_k$.
		\State $p_k \gets \frac{p_{\rm max}}{\widetilde{p}}p_k$, $\quad$ $k=1,\ldots,K.$
		\EndWhile
		\State {\bf Output:} $p_1,\ldots,p_K$.
	\end{algorithmic}
\end{algorithm}

As outlined in Algorithm~1, the steps of the fixed-point algorithm are in closed-form and computationally cheap. It has also been observed that the algorithm quickly converges to the optimal solution in tens of iterations.

\section{Computing Spatial Correlation Matrices with Local Scattering} \label{sec:derivation-spatial-correlation}

Practical RISs are expected to be deployed as planar arrays on walls and other building structures. The propagation channels will depend on both the azimuth and elevation angles of the impinging and departing radio waves.
To properly model the spatial fading correlation in such scenarios, we can utilize the fact that the normalized spatial correlation matrix $\vect{R}$ for any channel $\vect{h}$ can be computed as \cite[Sec.~7.3]{massivemimobook} 
\begin{equation} \label{eq:spatial-correlation}
\vect{R} = \iint f(\overline{\varphi},\overline{\theta}) \vect{a}(\overline{\varphi},\overline{\theta}) \vect{a}^{\Htran}(\overline{\varphi},\overline{\theta}) d\overline{\theta} d\overline{\varphi}
\end{equation}
 where the spatial scattering function $f(\overline{\varphi},\overline{\theta})$ describes the angular distribution of multipath components and satisfies $\iint f(\overline{\varphi},\overline{\theta}) d\overline{\theta} d\overline{\varphi}  = 1$. The array response vector captures the array geometry and is denoted by $\vect{a}(\overline{\varphi},\overline{\theta})$, where $\overline{\varphi}$ is the azimuth angle and $\overline{\theta}$ is the elevation angle. We consider a UPA deployed horizontally and vertically along the $y$ and $z$ axis, respectively. Taking the first antenna as the reference element, the $m$th element of $\vect{a}(\varphi,\theta)$ is given as
 \begin{align} 
\left[\vect{a}(\overline{\varphi},\overline{\theta})\right]_{m} = 
 e^{\imagunit2\pi d_{{\rm H}}^{m1}\sin(\overline{\varphi})\cos(\overline{\theta})}e^{\imagunit2\pi d_{{\rm V}}^{m1}\sin(\overline{\theta})}
\end{align}
where $d_{{\rm H}}^{m1}$ and $d_{{\rm V}}^{m1}$ denote the horizontal and vertical distances between the $m$th and the first array elements in wavelengths, respectively. In this paper, we consider the Gaussian local scattering model
 \cite[Sec.~2.6]{massivemimobook} where the scattering components are distributed around the nominal azimuth and elevation angles $\varphi$ and $\theta$, respectively,  according to a Gaussian distribution. The azimuth and elevation angular deviations are $\delta~\sim\mathcal{N}(0,\sigma_{\varphi}^2)$ and $\epsilon\sim\mathcal{N}(0,\sigma_{\theta}^2)$ independent of each other. Hence, the $(m,l)$th element of the normalized spatial correlation matrix $\vect{R}$ is 
\begin{align} \label{eq:ml-R}
\left[\vect{R}\right]_{ml} = 
\int\int & e^{\imagunit2\pi d_{{\rm H}}^{ml}\sin(\varphi+\delta)\cos(\theta+\epsilon)}e^{\imagunit2\pi d_{{\rm V}}^{ml}\sin(\theta+\epsilon)} \nonumber \\
&\times\frac{1}{2\pi\sigma_{\varphi}\sigma_{\theta}}e^{-\frac{\delta^2}{2\sigma_{\varphi}^2}}e^{-\frac{\epsilon^2}{2\sigma_{\theta}^2}}d\delta d\epsilon.
\end{align}
This expression can be evaluated numerically but the complexity can be high when we consider large arrays and RISs. In \cite[Sec.~2.6]{massivemimobook}, a closed-form approximate spatial correlation matrix is provided for uniform linear arrays (ULAs) in  two-dimensional world with only azimuth angular deviations.
We will generalize this result to UPAs in a three-dimensional world.

\begin{lemma}\label{lemma4} For the adopted UPA geometry and sufficiently small Gaussian angular deviations, the $(m,l)$th element of the normalized spatial correlation matrix in \eqref{eq:ml-R} is approximated as
\begin{align}
\left[\vect{R}\right]_{ml} 
\approx&\frac{ A_{ml} \widetilde{\sigma}_{ml}}{\sigma_{\varphi}} e^{\frac{D_{ml}^2\sigma_{\theta}^2\left(C_{ml}^2\sigma_{\theta}^2\widetilde{\sigma}_{ml}^2-1\right)}{2}}  \nonumber \\
&\times e^{-\imagunit B_{ml}C_{ml}D_{ml}\sigma_{\theta}^2\widetilde{\sigma}_{ml}^2}e^{-\frac{B_{ml}^2\widetilde{\sigma}_{ml}^2}{2}} \label{eq:spatial_correlation_UPA}
\end{align}
where 
\begin{align}
A_{ml}=& e^{\imagunit2\pi d_{{\rm H}}^{ml}\sin(\varphi)\cos(\theta)}e^{\imagunit2\pi d_{{\rm V}}^{ml}\sin(\theta)}, \nonumber\\  B_{ml}=&2\pi  d_{{\rm H}}^{ml}\cos(\varphi)\cos(\theta), \\
C_{ml}=&-2\pi d_{{\rm H}}^{ml}\cos(\varphi)\sin(\theta), \nonumber\\
D_{ml}=&-2\pi d_{{\rm H}}^{ml}\sin(\varphi)\sin(\theta)+ 2\pi d_{{\rm V}}^{ml}\cos(\theta), \\
\widetilde{\sigma}_{ml}^2=&\frac{\sigma_{\varphi}^2}{1+C_{ml}^2\sigma_{\varphi}^2\sigma_{\theta}^2}. \label{eq:sigmatilde}
\end{align}
\begin{proof}
	The proof is given in Appendix~\ref{appendix2}.
\end{proof}
\end{lemma}

Note that the derived approximate spatial correlation matrices can be computed easily for large-dimensional arrays. Moreover, it can also be utilized for ULAs in three-dimensional worlds with both azimuth and elevation angle deviations at the BS. We have observed that the SE values obtained with the approximate correlation matrices match very well with those obtained with exact spatial correlation matrices for $\sigma_{\varphi}=\sigma_{\theta} \leq 15^{\circ}$.

\section{Numerical Results and Discussion} \label{sec:numerical}

In this section, we compare the performance of the proposed RIS-assisted massive MIMO with several channel estimation and phase-shift selection schemes and conventional massive MIMO. 
The pathloss models and the shadow fading parameters for the LOS and NLOS paths originate from \cite[Table~5.1]{channel} for an urban microcell environment. 
Isotropic antennas are considered at the BS and UEs. The RIS elements have area $(\lambda/4)^2$, where $\lambda$ is the wavelength, and are deployed with $\lambda/4$ spacing. Unless otherwise stated, the carrier frequency is $1.9$\,GHz, the bandwidth is $1$\,MHz, and the thermal noise variance is $-107$\,dBm,  corresponding to a noise figure of $7$\,dB. The number of BS antennas is $M=100$ and they are deployed as a half-wavelength-spaced ULA. There are $L=2$ RISs, each being a UPA with $16\times16$ elements. The spatial correlation matrices for both the BS and RISs are generated using the closed-form formula for three-dimensional Gaussian local scattering model from Lemma~\ref{lemma4} with  15 degrees angular spread (i.e., $\sigma_{\varphi}=\sigma_{\theta}=\pi/12$). Each RIS is assigned to the one of the two UEs with the lowest BS-UE channel gains, by prioritizing the UE with the worst BS-UE channel and assigning the RIS with stronger channel to it. The coherence block length is $\tau_c=10\,000$ samples, which can be obtained for 1\,MHz coherence bandwidth (corresponding to the maximum difference of 300\,m between different paths) with 10\,ms coherence time (which corresponds to UE mobility of 7.89\,m/s, being equivalent to 28.4\,km/h) \cite[Sec.~2.1]{Marzetta2016a}.  Hence,  $\tau_c=10\,000$ is a realistic number for semi-static UEs as well as UEs that walk or run. For each scenario, the best integer $\tilde{\tau}$ is selected to have $K$ mutually orthogonal pilot sequences of length $\tau_p=\tilde{\tau}K$ for conventional massive MIMO by searching over the integers and picking the best one that results in the maximum average SE. For the RIS-assisted massive MIMO case, each $4\times4$ set of RIS elements is reconfigured to have the same phase-shift during pilot transmission for the short-term RIS configuration in Section~\ref{sec:individual} and, hence, we have $\tau_p=(LR+1)K=33K$. For the  channel estimation scheme with the long-term RIS configuration in Section~\ref{sec:overall}, $\dot{\tau}=33$ is selected to have the same pilot length, i.e., $\tau_p=\dot{\tau}K=33K$. 

\begin{figure*}[t]
\centering
		\begin{overpic}[width=8.6cm,tics=10]{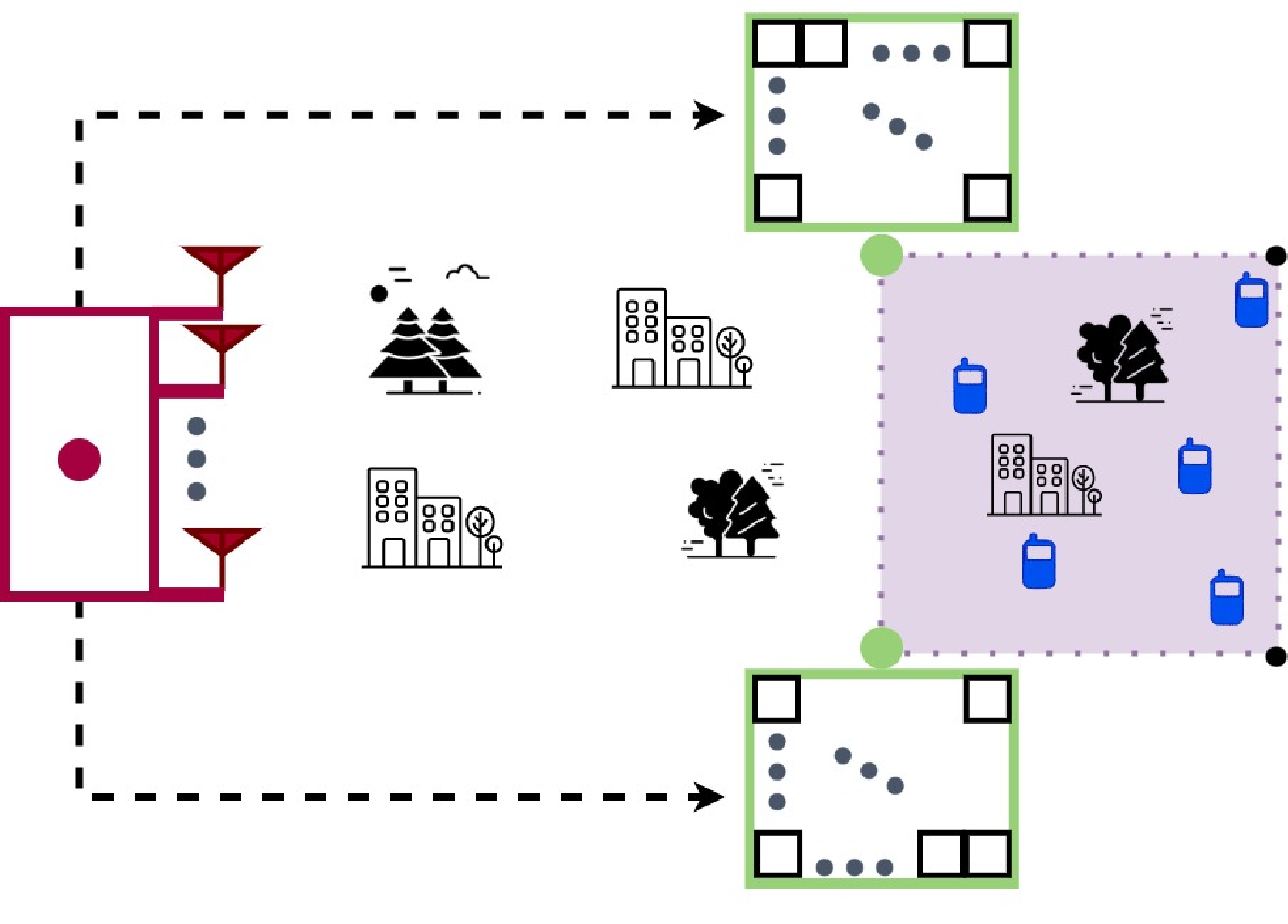}
			\put(-6,35.5){\small BS}
			\put(64,55){\small RIS 1}
			\put(64,16){\small RIS 2}
			\put(59.1,67.1){\footnotesize  1}
		\put(63,67.1){\footnotesize  2}	\put(75,55.1){\footnotesize  $N$}
		\put(75.9,4.2){\footnotesize  1}
		\put(71.9,4.2){\footnotesize  2}	\put(58.7,16.15){\footnotesize  $N$}
		\put(2,39){\small(0,0)}
			\put(20,49){\small 1}
		\put(20,42){\small 2}
			\put(19,27){\small $M$}
		\put(51,50){\small(200,50)}
		\put(51,21){\small(200,-50)}
	\put(100,50){\small(300,50)}
		\put(100,21){\small(300,-50)}
		\end{overpic}
	\caption{Experimental setup for the RIS-assisted massive MIMO for a single BS and two RISs. UEs are located in the colored region (cell-edge). }
	\label{fig:simulation-setup}
\end{figure*}

We assume the RISs are deployed to always have LOS paths to the BS. The existence of the LOS for the other channels is modeled in a probabilistic manner and the formulas for it and the Rician $\mathcal{K}$-factor are given in \cite[Sec.~5.5-3]{channel}, unless otherwise stated. The BS and the RISs are assumed to be mounted $10$\,m above the height of the UEs. The experimental setup is shown in Fig.~\ref{fig:simulation-setup}. The two-dimensional locations of the BS and two RISs is $(0,0)$, $(200,50)$, and $(200,-50)$, respectively, where the unit is meters. The antennas of the ULA at the BS are deployed horizontally along the $y$-axis while the UPAs at the RISs lie along the $x$- and $z$-axis. In each setup, $K=8$ cell-edge UEs are randomly dropped in the $100$\,m $\times\ 100\,$m area that extends from $(200,-50)$ to $(300,50)$. The maximum uplink power for each UE is $100$\,mW and each UE transmits with this power during pilot transmission unless otherwise stated. In the data transmission phase, the proposed max-min fair power control is adopted. In the figures, we plot the cumulative distribution function (CDF) of the SE per UE for conventional massive MIMO (Conv-mMIMO) and the proposed RIS-assisted massive MIMO (RIS-mMIMO) with different receive combining schemes. The randomness is the UE locations and shadowing.

\subsection{Simulation Results}

\begin{figure}[h!]
			\includegraphics[trim={0.2cm 0.1cm 0.2cm 0.8cm},clip,width=3.7in]{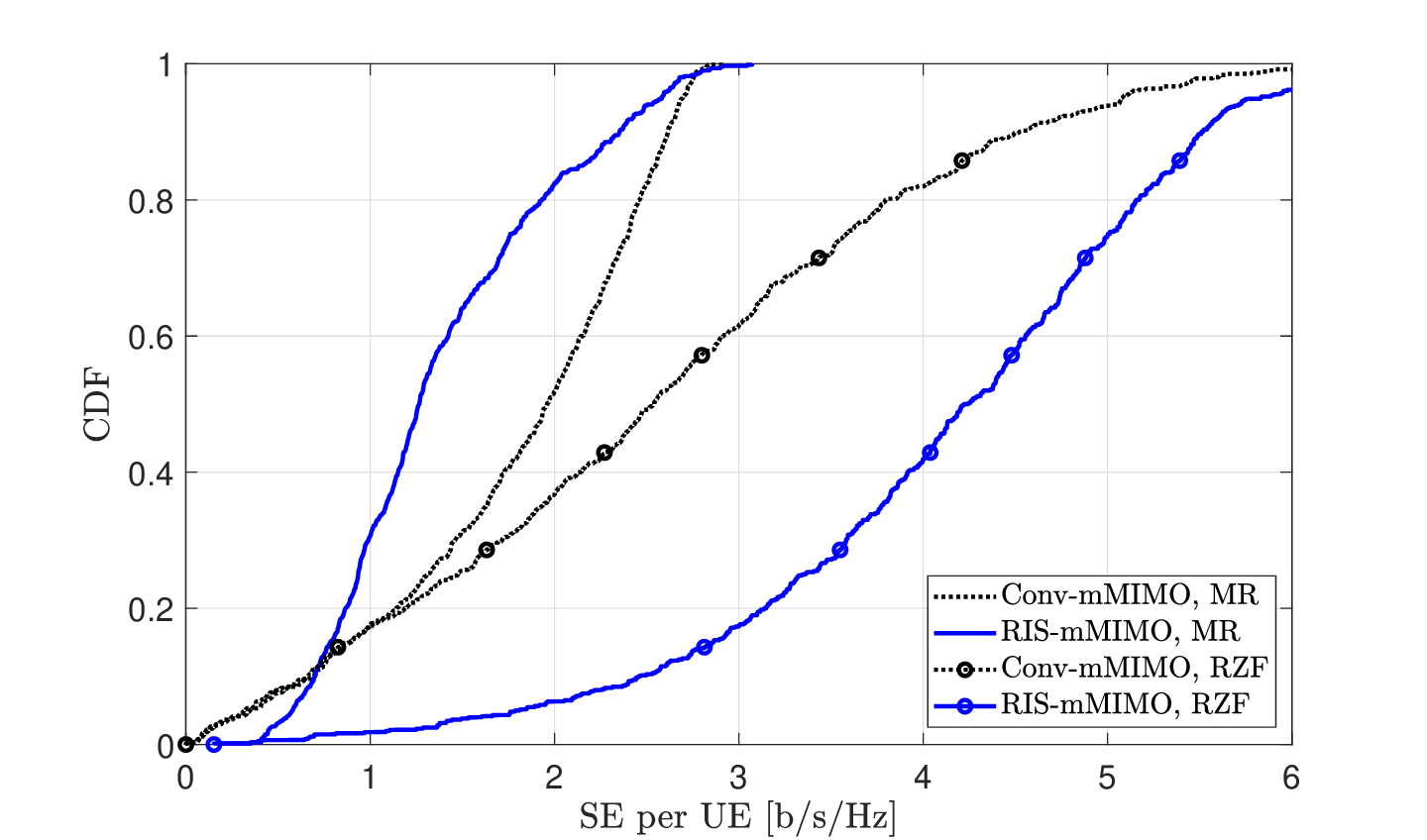}
			\caption{CDF of SE per UE for conventional and RIS-assisted massive MIMO with different receive combiners. } \label{fig:fig3}
		\end{figure}

In Fig.~\ref{fig:fig3}, Conv-mMIMO and RIS-mMIMO are considered when using either MR or RZF receive combining. For the RIS-mMIMO, the proposed channel estimation for the short-term RIS configuration (the first scheme in Fig.~\ref{fig:channel_estimation_schemes}) and the most advanced phase-shift selection scheme in \eqref{eq:phase-shift1} are considered. For all the channels, the LOS path is the only specular component once an LOS exists. For MR combining, RIS-mMIMO starts from a higher SE in the lower tail of the CDF curve than with Conv-mMIMO, which indicates that the RISs effectively improves the SNRs and this benefits the most unfortunate UEs. To quantify this, the 95\%-likely SE (the point where the CDF is 0.05) can be observed and it is approximately 1.8 times greater with RIS-mMIMO. However, this comes with a significant SE reduction for other UEs. The reason is that the RIS creates strong additional  channel components compared to the non-RIS case.
This results in extra co-user interference and the MR combining scheme is not designed to manage that.
However,  RIS-mMIMO provides a significant SE improvement for all UEs when RZF combining is used to suppress interference. The median SE (the point where the CDF is 0.5) provided by RIS-mMIMO is 1.7 times larger compared to Conv-mMIMO with RZF combining. Moreover, the 90\%-likely SE (the point where the CDF is 0.1) is approximately 3.7 times greater with RIS-mMIMO. The lower tails of the CDF curves are of great importance since low-rate UEs need to be active more often than high-rate UEs, when using the same service.

\begin{figure}[h!]
			\includegraphics[trim={0.2cm 0.1cm 0.2cm 0.8cm},clip,width=3.7in]{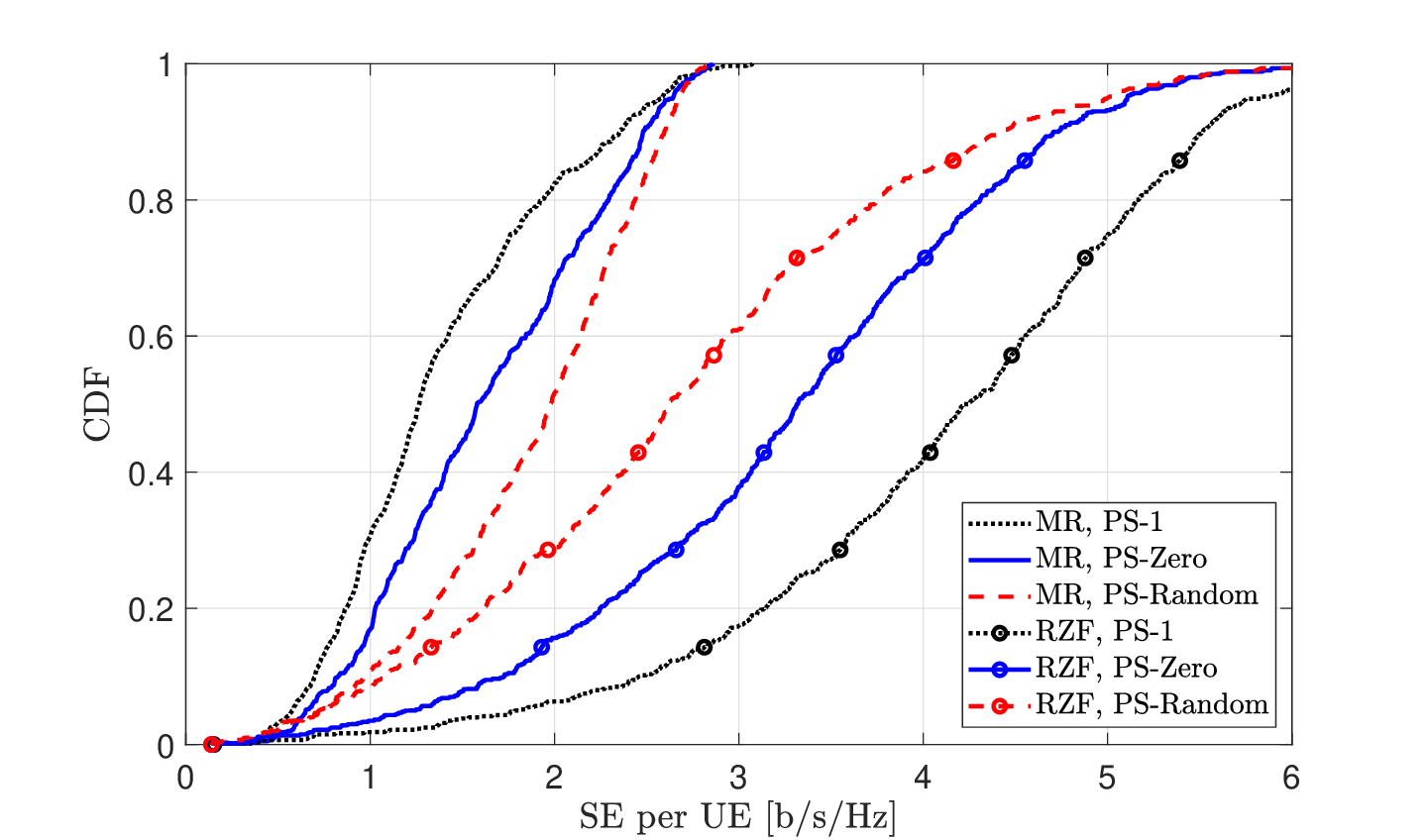}
			\caption{CDF of SE per UE for RIS-assisted massive MIMO with different phase-shift configurations. } \label{fig:fig4}
\end{figure}

In Fig.~\ref{fig:fig4}, we reconsider the MR and RZF curves from Fig.~\ref{fig:fig3} for RIS-mMIMO case, which are labeled as ``PS-1'' corresponding the first phase-shift selection scheme in \eqref{eq:phase-shift1}. For this specific scenario, the phase-shift selection scheme in  \eqref{eq:phase-shift3} is omitted since its performance is almost identical. In addition to PS-1, two other phase-shift selection schemes are considered. The first one is ``PS-Zero'' where all the RIS elements are configured to have zero phase-shift. The second scheme is ``PS-Random'' where random RIS phase-shifts are considered irrespective of the channel estimates. For RZF combining, there is a significant performance drop compared to the proposed phase-shift selection scheme PS-1. When we compare this figure with Fig.~\ref{fig:fig3}, it is seen that both PS-Zero and PS-Random improve the SE compared to Conv-mMIMO, but not as great as the well-configured RIS with PS-1. This improvement comes from the SNR increase due to the newly added channel components compared to the original Conv-mMIMO environment. 
Interestingly, when MR combining is used, it is better to use zero or random phase-shifts than using PS-1. As discussed in relation to Fig.~\ref{fig:fig3}, this is due to the extra interference that an optimized RIS can cause.
In conclusion, a well-optimized RIS configuration requires an interference-suppressing receive combining scheme to be practically useful.


From now on, we will only consider RZF combining for RIS-mMIMO and Conv-mMIMO, respectively. We consider the same scenario as in Figs.~\ref{fig:fig3} and \ref{fig:fig4} to compare the proposed LMMSE-based channel estimators with the element-wise (EW) LMMSE and LS-based estimators. The EW LMMSE estimator is obtained by neglecting the off-diagonal entries of the spatial correlation matrices in \eqref{eq:hk_hat} and \eqref{eq:f_kell_estimates} for short-term configuration and in \eqref{eq:h_kdot_estimates} for long-term configuration, respectively. It is a low-complexity alternative of LMMSE estimator and only requires the knowledge of channel gains \cite[Sec.~3.4.1]{massivemimobook}. LS estimator does not utilize any channel statistics and is applied from \cite{RISchannelEstimation_nested_cascaded, RISchannelEstimation_nested_LS_single}  to our problem setup in this paper. We consider the EW MMSE estimator for the long-term RIS configuration when using random phase shifts. This scheme is the simplest in the sense that the spatial correlation is used neither for the channel estimation nor for the RIS configuration. We replot the curve corresponding to the proposed LMMSE-based channel estimation for the short-term RIS configuration.

\begin{figure}[h!]
			\includegraphics[trim={0.2cm 0.1cm 0.2cm 0.8cm},clip,width=3.7in]{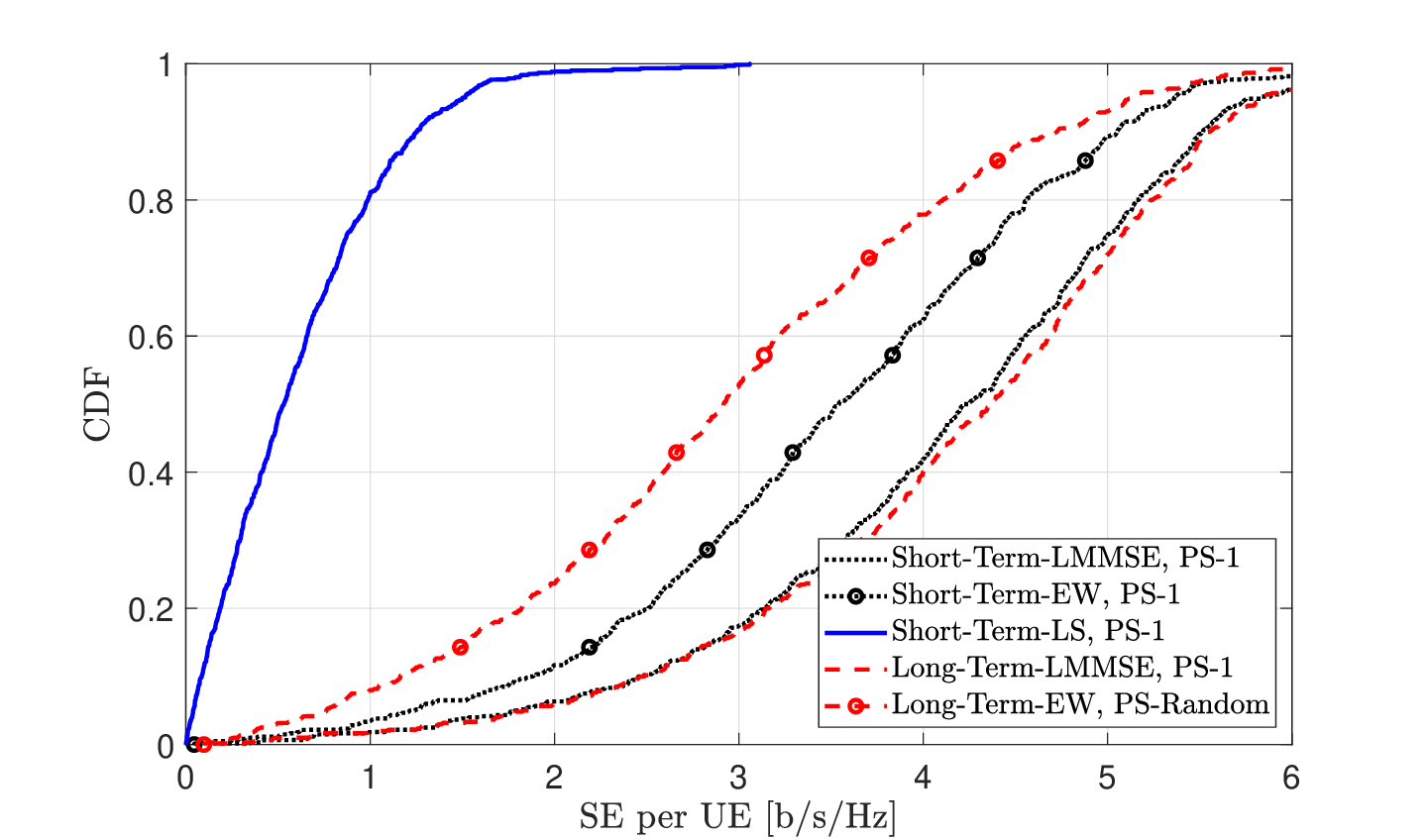}
			\caption{CDF of SE per UE for different  channel estimators and RIS configurations.  } \label{fig:fig5}
		\end{figure}

As shown in Fig.~\ref{fig:fig5}, there is a significant SE reduction when the LS estimator is used for the short-term RIS configuration. Recall that the individual paths are estimated with the RIS subsets of elements that share the same phase-shift during pilot transmission in this configuration. When we do not take into account the channel correlation and the LOS paths, the channel estimation quality of the LS estimator is inferior due to the resulting pilot contamination created by the multiple RIS elements that share the same phase-shifts. When the EW-LMMSE estimator is utilized, there is a substantial improvement in the SE. However, using the proposed LMMSE estimator that exploits the spatial correlation performs much better. On the other hand, we have observed that both the EW LMMSE and LS estimators for the long-term RIS configuration with PS-1 are almost identical to their LMMSE counterpart since only the overall channels are estimated with higher accuracy. For simplicity, we haven't included the corresponding curves in the figure. In this case, exploiting channel statistics for channel estimation does not bring additional advantage. However, selecting the RIS phase-shifts randomly results in a significant performance degradation. Hence, using spatial correlation is of critical importance when it comes to RIS reconfiguration. As a last observation, note that the performance of the long-term RIS configuration is slightly better than that of the short-term RIS configuration. Since the latter has a more complicated structure and requires changing the RIS configuration in every coherence block, for this scenario with only one specular component, the long-term RIS configuration by using the dominant eigenvectors of the spatial correlation matrices is preferable. In the following part, we will explore other scenarios to gain more insights.

\begin{figure}[h!]
			\includegraphics[trim={0.2cm 0.1cm 0.2cm 0.8cm},clip,width=3.7in]{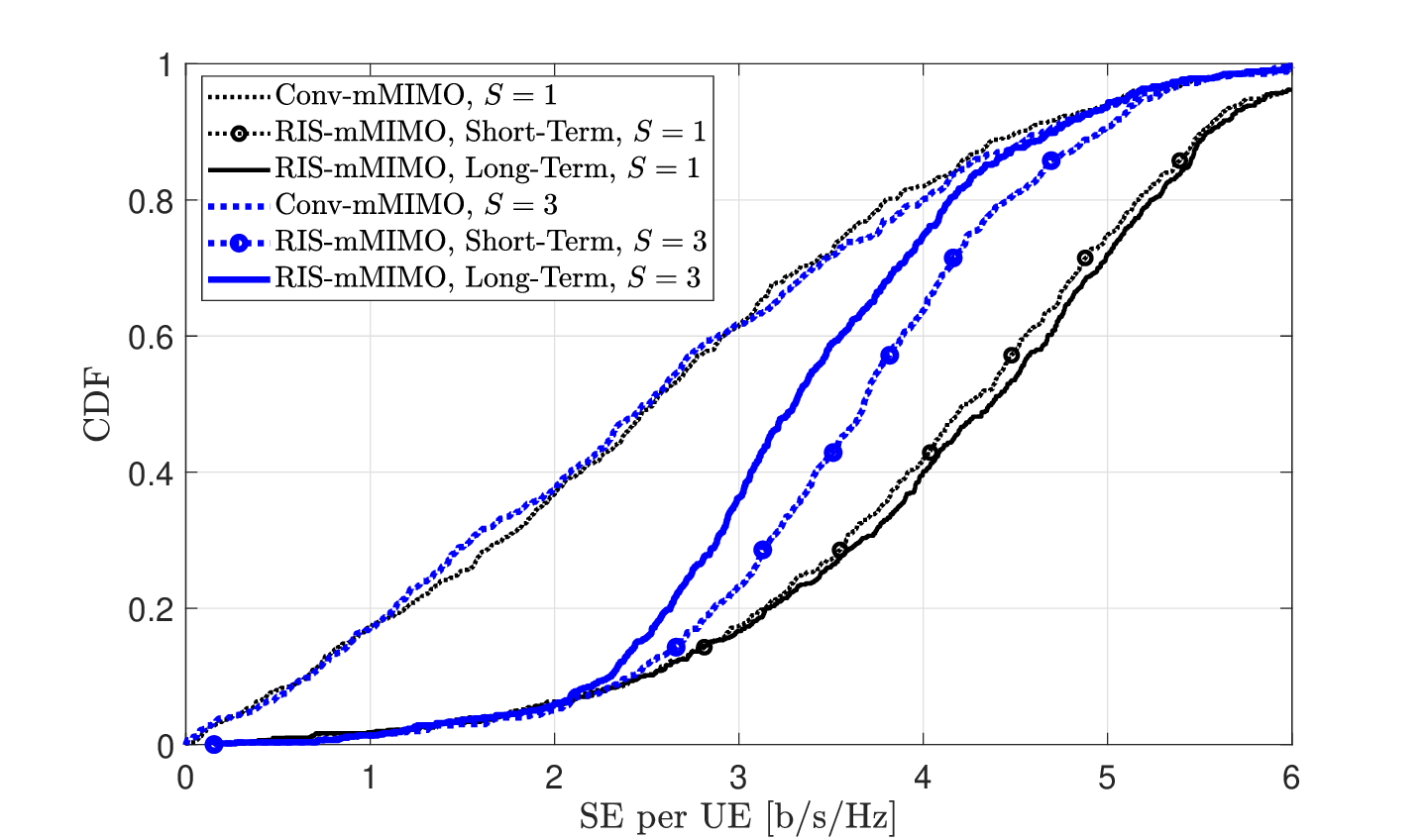}
			\caption{CDF of SE per UE for different number of specular components.} \label{fig:fig6}
\end{figure}

In the remaining simulations, we consider LMMSE channel estimators and the proposed phase-shift selection method. Until now, we have assumed that LOS path is the only specular component once LOS exists. In Fig.~\ref{fig:fig6}, we compare this scenario with the case of $S=3$ specular components for all the channels. In this case, the original LOS channel gain (once exists) is distributed randomly over two non-LOS dominant components by keeping the power ratio of the LOS component being 0.5. Note that the angles-of-arrival and angles-of-departure for the non-LOS specular components are generated randomly around the LOS angles within 60 degrees and 15 degrees for the azimuth and elevation angles, respectively. Although the power is distributed among three specular components, RIS-mMIMO provides much higher SE in particular for the unfortunate UEs. With more specular components, while the SE performance of the conventional massive MIMO does not change considerably, there is a loss in the upper tails of the CDF curves for the RIS-assisted massive MIMO. Since the power of the LOS component is lower, most of the time, there is a SE reduction. We also observe that when there is only one dominant component, the long-term RIS configuration is slightly better. However, when having the same channel gain distributed over three specular components, the short-term RIS configuration results in better performance with a $11\%$ improvement in the median SE.


\begin{figure}[h!]
			\includegraphics[trim={0.2cm 0.1cm 0.2cm 0.8cm},clip,width=3.7in]{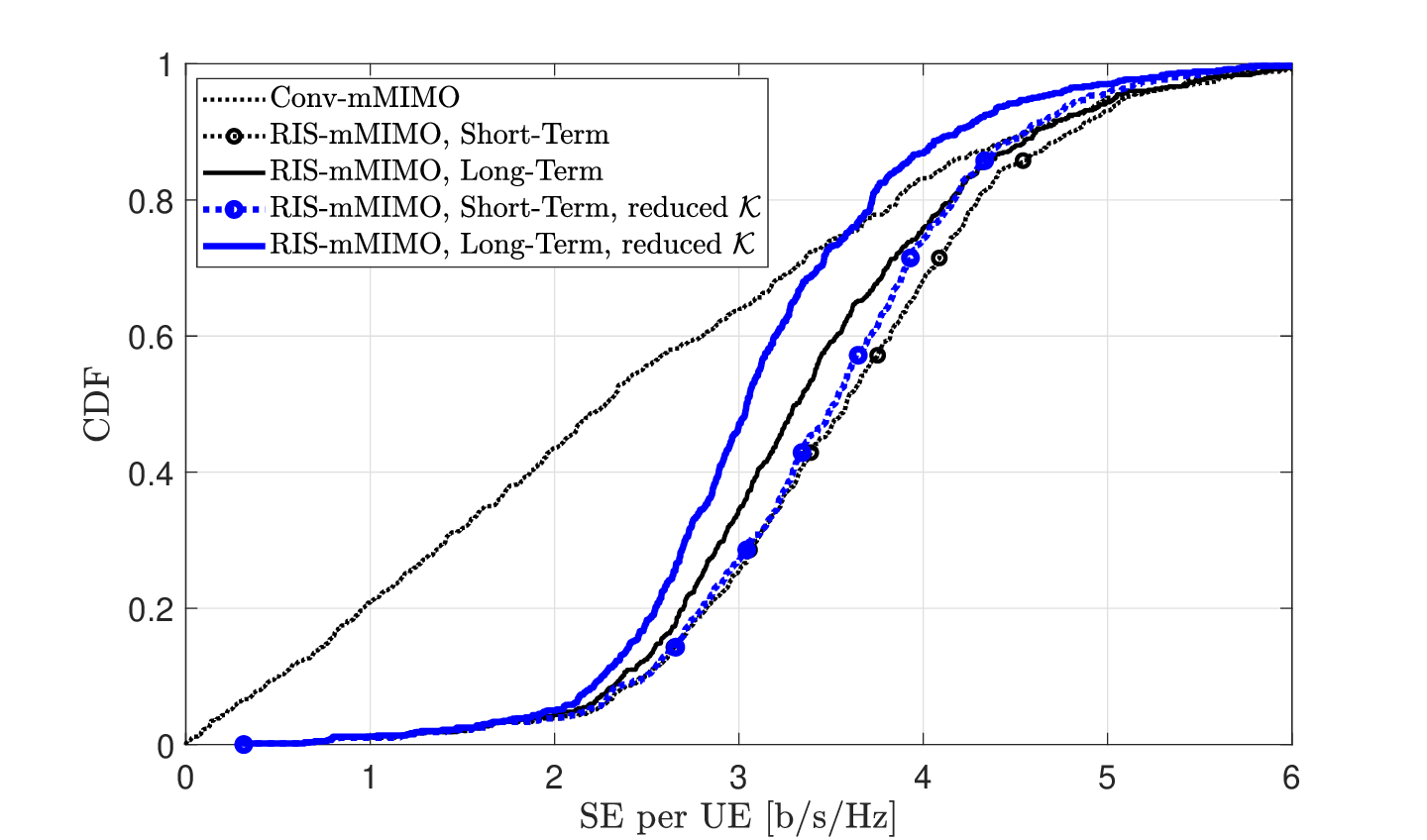}
			\caption{CDF of SE per UE  when the number of specular components of the  channels is five. There is either a 9\,dB reduction in Rician $\mathcal{K}$-factors or not.} \label{fig:fig7}
		\end{figure}

To see the effect of the Rician factor on the proposed RIS-mMIMO schemes, we consider the same scenario from Fig.~\ref{fig:fig6} with $S=5$ specular components. In addition, we consider the scenario with $S=5$ specular components in which a 9\,dB reduction is introduced on the distance-dependent Rician $\mathcal{K}$-factors of the RIS-UE channels in \cite[Sec.~5.5-3]{channel}. The main reason to consider such a scenario is to keep the high channel gain in the LOS case from the model in \cite{channel} but redistribute so that the specular components are less dominant compared to the nonspecular Rayleigh fading. As in the case of three specular components, the short-term RIS configuration provides greater SE in comparison to the long-term RIS configuration. When there is reduction in the Rician $\mathcal{K}$-factors, short-term RIS configuration based on small-scale channel estimates  gains more importance than the overall channel estimation with a long-term RIS phase-shift configuration. As shown in Fig.~\ref{fig:fig7}, it is seen that the CDF curve of short-term RIS configuration becomes more rightward than that of overall channel estimation compared to the original case where there is no reduction in the original Rician $\mathcal{K}$-factor.

\begin{figure}[h!]
			\includegraphics[trim={0.2cm 0.1cm 0.2cm 0.8cm},clip,width=3.7in]{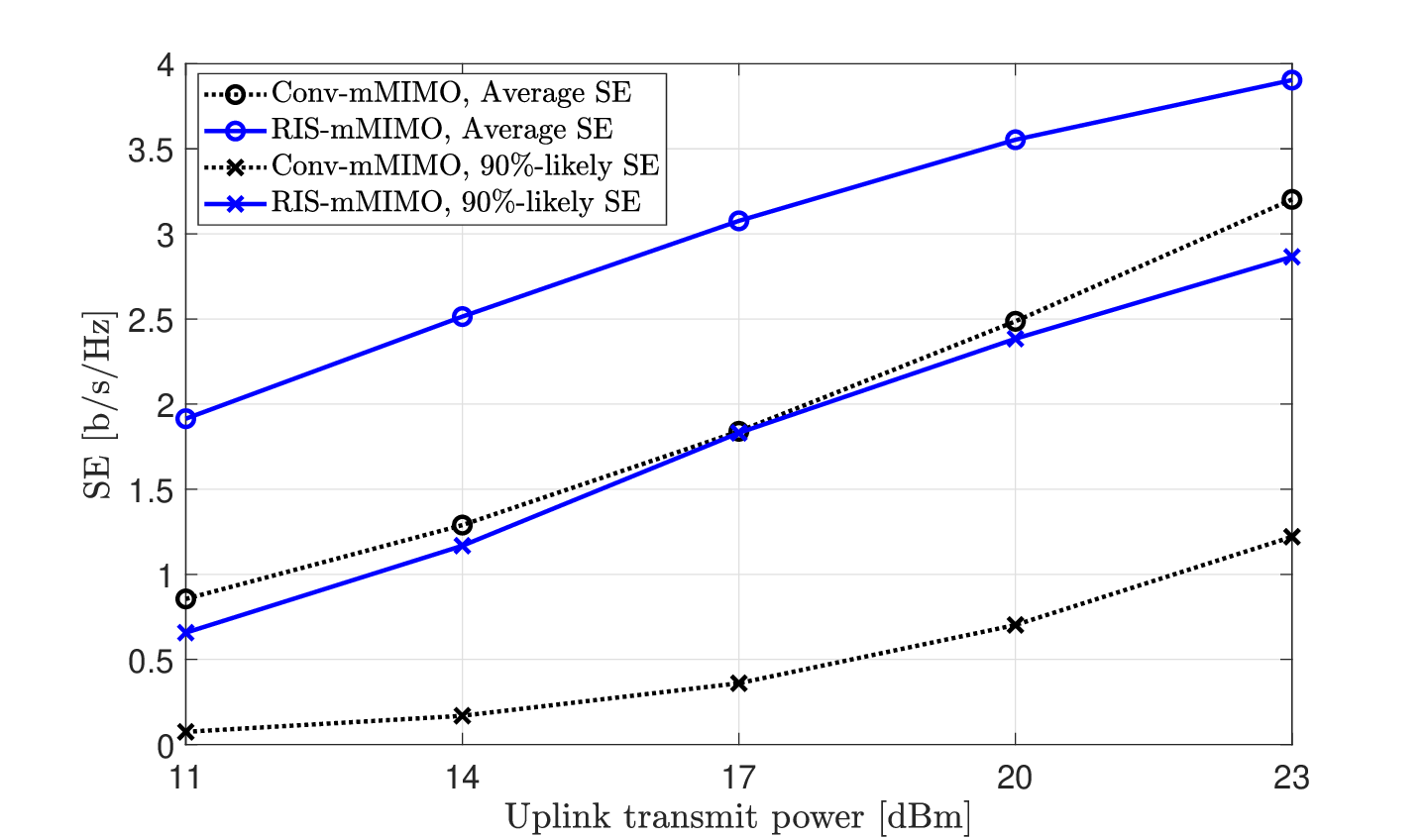}
			\caption{Average and $90\%$-likely SE for conventional and RIS-assisted massive MIMO with short-term configuration.} \label{fig:fig8}
\end{figure}

In Fig.~\ref{fig:fig8}, the average and $90\%$-likely SE for the conventional  and RIS-assisted massive MIMO with short-term configuration are plotted while changing the uplink transmit power. There are $S=5$ specular components. RIS assistance benefits both the average SE and the $90\%$-likely SE. Regarding the percentage improvement, it is seen that RISs are more helpful at low SNR since the SNR increase provided by RIS assistance becomes more effective.

\begin{figure}[h!]
			\includegraphics[trim={0.2cm 0.1cm 0.2cm 0.8cm},clip,width=3.7in]{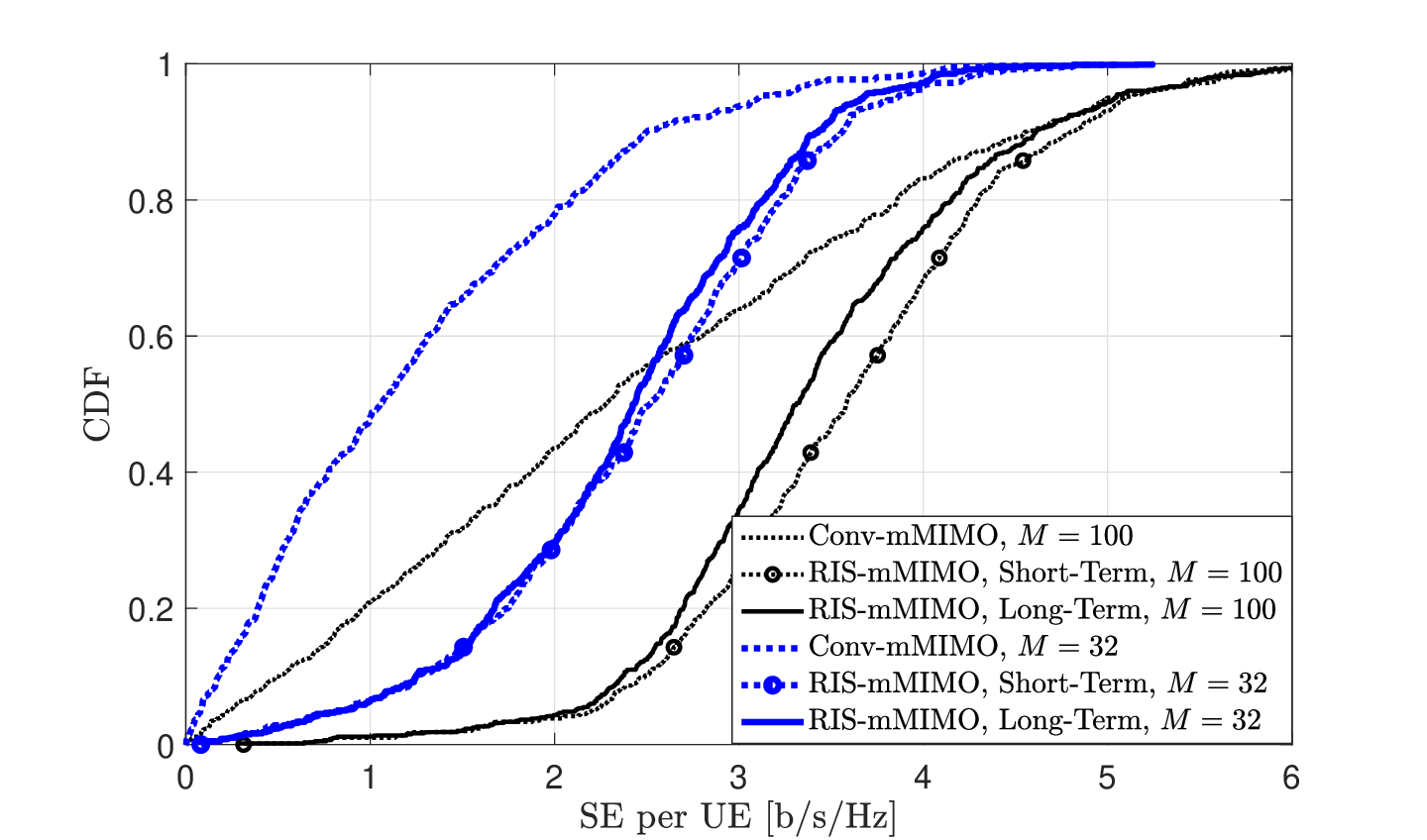}
			\caption{CDF of SE per UE for different number of BS antennas.} \label{fig:fig9}
		\end{figure}

In Fig.~\ref{fig:fig9}, we consider the same setup as in Fig.~\ref{fig:fig8} by setting the uplink transmit power as $100$\,mW ($20$\,dBm). We consider two cases: i) $M=100$ and ii) $M=32$ BS antennas to see the impact of RIS assistance when having smaller number of antennas. As expected, all the curves move leftward when having $M=32$. For smaller number of antennas, RIS-assisted massive MIMO still outperforms the conventional massive MIMO significantly. Another important result is that the additional SE increase provided by the short-term RIS configuration becomes less dominant when having $M=32$. Hence, we conclude that channel estimation is more preferable in selecting the RIS phase-shifts at small-scale level when the number of antennas is greater. 

\begin{figure}[h!]
			\includegraphics[trim={0.2cm 0.1cm 0.2cm 0.8cm},clip,width=3.7in]{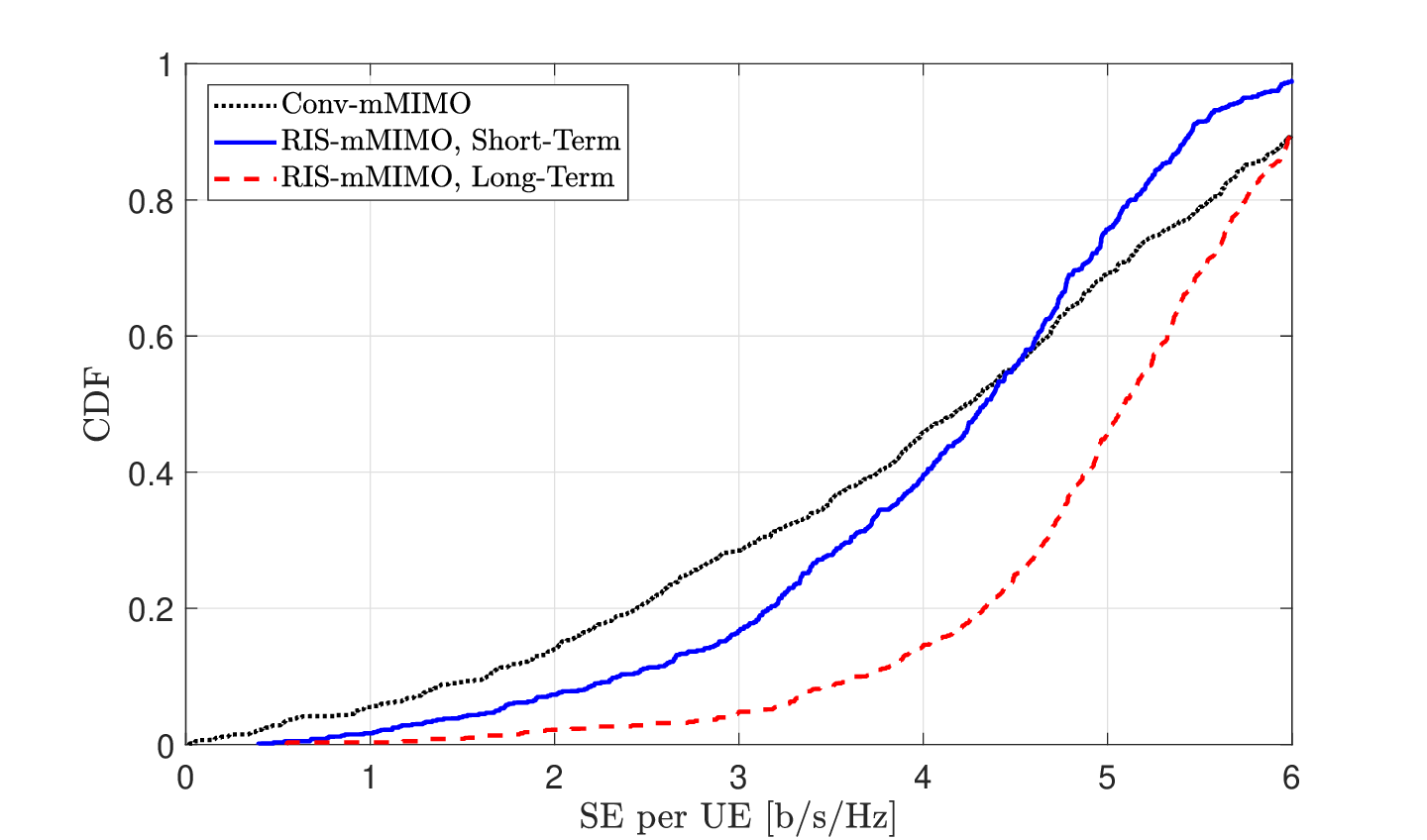}
			\caption{CDF of SE per UE for $\tau_c=1000$.} \label{fig:fig10}
\end{figure}

Finally, in Fig.~\ref{fig:fig10}, we consider a smaller coherence block length compared to the previous case. Now, the communication bandwidth (also coherence bandwidth) and the number of samples in each coherence block are reduced by ten, i.e., we consider 100\,kHz bandwidth and $\tau_c=1000$. Due to the reduced number of samples, we also consider less number of pilots for the RIS-assisted massive MIMO. Each $8\times8$ set of RIS elements is reconfigured to have the same phase-shift during pilot transmission for the short-term RIS configuration and, hence, we have $\tau_p=(LR+1)K=9K$. Maximum uplink transmit power is $50$\,mW and the LOS path is the only specular component once it exists. For the long-term RIS configuration the same pilot length is considered. It is worth mentioning that for the conventional massive MIMO, we still consider the best pilot length that results in the maximum average SE. As Fig.~\ref{fig:fig10} demonstrates, the long-term RIS configuration provides greater SE than conventional massive MIMO, in particular in the unfortunate cases. Since a greater number of RIS elements share the same phase-shift during pilot transmission for the short-term configuration, resolving individual paths becomes more challenging and the SE decreases compared to the long-term RIS configuration. However, both RIS configurations outperform the conventional massive MIMO. We have observed that using channel estimation error statistics can be useful in boosting the SE with the short-term configuration, for which we haven't provided any results in this paper, still the long-term configuration is better than the short-term configuration when the channel coherence length limits the number of resolvable paths during channel estimation.

\section{Conclusion} \label{sec:conclusion}

When RISs are used to support Massive MIMO, must the phase-shift be selected based on estimates of the small-scale channel fading or is sufficient to consider the long-term statistics? To answer this question, we have proposed several LMMSE estimation schemes with low training overhead and the respective phase-shift and receive combiner design. We consider 
a multi-specular spatially correlated fading environment. Configuring the RIS based on the long-term  spatial correlation statistics without using the channel estimates and by estimating only the overall channel is less complicated compared to short-term RIS reconfiguration based on estimating the fading variations of the individual paths. When the channel coherence block is sufficiently large and there are multiple specular channel components, estimating the individual RIS paths and using those estimates in RIS configuration outperforms the long-term RIS configuration.  This is due to the fact that the dominant eigenvalues of the spatial correlation matrices, which are used in the long-term configuration, do not capture the channel characteristics sufficiently well. In addition, when the Rician-$\mathcal{K}$ factor of the RIS-UE channels is smaller, the SE gain obtained by the short-term configuration increases. This is a natural consequence of the weaker specular components, and thus the weaker dominant eigenvectors of the spatial correlation matrices. In this case, exploiting the small-scale part of the channel in RIS configuration gains importance.

On the other hand, when the LOS component is the only specular component once it exists, the long-term configuration slightly outperforms the short-term configuration. Since the former RIS configuration has also less signaling overhead, it is preferable in this scenario. Moreover, when the channel coherence block has relatively less number of samples, long-term configuration is more useful. This is because stricter pilot length requirement of the short-term configuration enforces construction of larger RIS subsets sharing the same phase-shifts during pilot transmission phase, which in turn leads to the estimation accuracy reduction of the individual RIS paths.

\appendices
\section{Proof of Lemma~\ref{lemma:norm_maximization}} \label{appendix0}

We first re-express the problem in \eqref{eq:optimization2} as follows:
\begin{subequations} \label{eq:optimization3}
\begin{align} 
&\underset{\bm{\phi}_j}{\textrm{maximize}} \quad 
\bm{\phi}_j^{\Htran}\vect{A}_j\bm{\phi}_j+2\Re\left(\bm{\phi}_j^{\Htran}\vect{b}_j\right)
 \label{eq:norm-maximization5} \\
&\textrm{subject to} \quad \Vert \bm{\phi}_j\Vert^2\leq N_j \label{eq:norm-maximization6}\end{align}
\end{subequations}
where $\vect{A}_j\triangleq(\widehat{\vect{H}}^{\prime}_{jj})^{\Htran}\widehat{\vect{H}}^{\prime}_{jj}$ and $\vect{b}_j\triangleq(\widehat{\vect{H}}^{\prime}_{jj})^{\Htran}\widehat{\vect{h}}_j$. To obtain a more tractable form of the problem, we first get the eigendecomposition of $\vect{A}_j$ as $\vect{A}_j=\sum_{d=1}^{N_j}\lambda_{j,d}\vect{u}_{j,d}\vect{u}_{j,d}^{\Htran}$ where $\lambda_{j,d}\geq0$ are the nonnegative eigenvalues and $\vect{u}_{j,d}\in \mathbb{C}^{N_j}$ are the orthonormal eigenvectors. We then express the main optimization variable as $\bm{\phi}_j=\sum_{d=1}^{N_j}\alpha_{j,d}e^{\imagunit\theta_{j,d}}\vect{u}_{j,d}$ with real $\alpha_{j,d}\geq0$. The problem in \eqref{eq:optimization3} becomes in terms of $\alpha_{j,d}$ and $\theta_{j,d}$, for $d=1,\ldots,N_j$, as follows:
\begin{subequations} \label{eq:optimization4}
 \begin{align} 
 &\underset{\alpha_{j,d}, \theta_{j,d}, \ d=1,\ldots,N_j}{\textrm{maximize}} \quad 
\sum_{d=1}^{N_j}\left(\lambda_{j,d}\alpha_{j,d}^2+2\alpha_{j,d}\Re\left(e^{-\imagunit\theta_{j,d}}\vect{u}_{j,d}^{\Htran}\vect{b}_j\right)\right)
 \label{eq:norm-maximization7} \\
 &\hspace{0.4cm}\textrm{subject to} \quad \hspace{0.4cm} \sum_{d=1}^{N_j}\alpha_{j,d}^2\leq N_j, \ \alpha_{j,d}\geq 0, \ d=1,\ldots,N_j. \label{eq:norm-maximization8}\end{align}
 \end{subequations}
When there is no direct link between the BS and UE $j$, we can set the corresponding channel estimate $\widehat{\vect{h}}_j$, and, hence $\vect{b}_j$ to zero. In this case, the problem is independent of $\theta_{j,d}$ and the optimal $\bm{\phi}_j$ becomes
\begin{equation}
\bm{\phi}_j^{\star}=\sqrt{N_j}\vect{u}_{j,\bar{d}}
\end{equation}
where $\bar{d}$ is the index corresponding to the dominant eigenvector of $\vect{A}_j$. For the other case where $\vect{b}_j\neq \vect{0}_{N_j}$,
 it can easily be shown that the optimal phase-shifts for the above problem are given by $\theta_{j,d}^{\star}=\angle\vect{u}_{j,d}^{\Htran}\vect{b}_j$. The Karush-Kuhn-Tucker conditions for the optimal solution of \eqref{eq:optimization4} after inserting these optimal phase-shifts to \eqref{eq:norm-maximization7} are given as
 \begin{align}
 &2\lambda_{j,d}\alpha_{j,d}+2\vert\vect{u}_{j,d}^{\Htran}\vect{b}_j\vert = 2\gamma_1\alpha_{j,d}-\gamma_{2,d}, \quad d=1,\ldots,N_j \label{eq:KKT1}\\
 &\gamma_1\left(\sum_{d=1}^{N_j}\alpha_{j,d}^2-N_j\right)=0 \label{eq:KKT2} \\
 &\gamma_{2,d}\alpha_{j,d}=0,\quad d=1,\ldots,N_j \label{eq:KKT3} \\
 &\gamma_1\geq 0, \quad \gamma_{2,d}\geq 0, \quad d=1,\ldots,N_j \label{eq:KKT4}
 \end{align}
 together with the constraints in \eqref{eq:norm-maximization8}. Here, $\gamma_1$ and $\gamma_{2,d}$ are the Lagrange multipliers corresponding to the inequality constraints in \eqref{eq:norm-maximization8}, respectively. We notice that $\alpha_{j,d}$ cannot be zero by \eqref{eq:KKT1} and \eqref{eq:KKT4} if $\vert\vect{u}_{j,d}^{\Htran}\vect{b}_j\vert\neq0$, which leads to $\gamma_{2,d}=0$ for those $d$ such that $\vert\vect{u}_{j,d}^{\Htran}\vect{b}_j\vert\neq0$ by \eqref{eq:KKT3}. If $\vert\vect{u}_{j,d}^{\Htran}\vect{b}_j\vert=0$, we should again have $\gamma_{2,d}=0$ by \eqref{eq:KKT1} and \eqref{eq:KKT3}. Hence, $\gamma_{2,d}=0$,  $\forall d$ . Furthermore, $\gamma_1$ cannot be zero by $\eqref{eq:KKT1}$ and we should have  $\sum_{d=1}^{N_j}\alpha_{j,d}^2=N_j$ from \eqref{eq:KKT2}. So, the optimal $\alpha_{j,d}$, for $d=1,\ldots,N_j$ from \eqref{eq:KKT1} should be
\begin{align}
\alpha_{j,d}^{\star}=\frac{\vert\vect{u}_{j,d}^{\Htran}\vect{b}_j\vert}{\gamma_1^{\star}-\lambda_{j,d}}, \quad d=1,\ldots,N_j \label{eq:alpha}
\end{align}
where $\gamma_1^{\star}> \max_{d}\lambda_{j,d}$ is the unique root of
\begin{align}
\sum_{d=1}^{N_j}\frac{\vert\vect{u}_{j,d}^{\Htran}\vect{b}_j\vert^2}{\left(\gamma_1-\lambda_{j,d}\right)^2}=N_j \label{eq:root}
\end{align}
which can easily be found by a simple bisection search since the left-hand side of \eqref{eq:root} is monotonically decreasing for $\gamma_1> \max_{d}\lambda_{j,d}$. Inserting $\alpha_{j,d}^{\star}$ in \eqref{eq:alpha} and $\theta_{j,d}^{\star}=\angle\vect{u}_{j,d}^{\Htran}\vect{b}_j$ to $\bm{\phi}_j=\sum_{d=1}^{N_j}\alpha_{j,d}e^{\imagunit\theta_{j,d}}\vect{u}_{j,d}$, we complete the proof.

\section{Proof of Lemma~\ref{lemma1}}
\label{appendix1}

Since $\left[\vect{H}_{k\ell}\right]_{:n}$ has zero mean, from estimation theory \cite{Kay1993a}, the matrices $\overline{\vect{R}}_{k\ell,n}$ and $\overline{\vect{R}}_{k(\ell,r)}$ are given by
\begin{align}
&\overline{\vect{R}}_{k\ell,n}=\mathbb{E}\left\{ \left[\vect{H}_{k\ell}\right]_{:n} \sum_{n^{\prime}\in \mathrm{RIS}_{\ell,r}} \left[\vect{H}_{k\ell}\right]_{:n^{\prime}}^{\Htran}\right\}, \nonumber \\
&\overline{\vect{R}}_{k(\ell,r)}=\mathbb{E}\left\{ \sum_{n\in \mathrm{RIS}_{\ell,r}} \left[\vect{H}_{k\ell}\right]_{:n} \sum_{n^{\prime}\in \mathrm{RIS}_{\ell,r}} \left[\vect{H}_{k\ell}\right]_{:n^{\prime}}^{\Htran}     \right\}. \label{eq:Rklr}
\end{align}
Defining $\boldsymbol{\mathcal{A}}_{k\ell,nn^{\prime}}\triangleq \mathbb{E}\left\{\left[\vect{H}_{k\ell}\right]_{:n}\left[\vect{H}_{k\ell}\right]_{:n^{\prime}}^{\Htran}\right\}$, we can express the correlation matrices in \eqref{eq:Rklr} as
\begin{equation}
\overline{\vect{R}}_{k\ell,n}=\sum_{n^{\prime}\in \mathrm{RIS}_{\ell,r}} \boldsymbol{\mathcal{A}}_{k\ell,nn^{\prime}},
\quad \overline{\vect{R}}_{k(\ell,r)}= \sum_{n\in \mathrm{RIS}_{\ell,r}}  \overline{\vect{R}}_{k\ell,n}. \label{eq:Rklr2}
\end{equation}
So, we need to obtain $\boldsymbol{\mathcal{A}}_{k\ell,nn^{\prime}}$, $\forall n, \forall n^{\prime}$. From \eqref{eq:cascaded2}, we have
\begin{equation}
\boldsymbol{\mathcal{A}}_{k\ell,nn^{\prime}}=\mathbb{E}\left\{ \left[\vect{G}_{\ell}\right]_{:n}\left[\vect{f}_{k\ell}\right]_n\left[\vect{f}_{k\ell}\right]_{n^{\prime}}^*\left[\vect{G}_{\ell}\right]_{:n^{\prime}}^{\Htran}\right\}
\end{equation}
where $\left[\vect{f}_{k\ell}\right]_n$ denotes the $n$th element of the vector $\vect{f}_{k\ell}$. Using the independence of the channels $\vect{G}_{\ell}$ and $\vect{f}_{k\ell}$, we have
\begin{equation}
\boldsymbol{\mathcal{A}}_{k\ell,nn^{\prime}}=\mathbb{E}\left\{\left[\vect{G}_{\ell}\right]_{:n}\left[\overline{\vect{R}}_{k\ell}^{\mathrm{f}}\right]_{nn^{\prime}}\left[\vect{G}_{\ell}\right]_{:n^{\prime}}^{\Htran}\right\} 
\end{equation}
where $\left[\overline{\vect{R}}_{k\ell}^{\mathrm{f}}\right]_{nn^{\prime}}$ denotes the $(n,n^{\prime})$th element of the correlation matrix $\overline{\vect{R}}_{k\ell}^{\mathrm{f}}$ that is given as
\begin{equation}
\overline{\vect{R}}_{k\ell}^{\mathrm{f}}=\mathbb{E}\left\{\vect{f}_{k\ell}\vect{f}_{k\ell}^{\Htran}\right\}=\sum_{s=1}^{S_{k\ell}^{\mathrm{f}}}\bar{\vect{f}}_{k\ell,s}\bar{\vect{f}}_{k\ell,s}^{\Htran}+\vect{R}_{k\ell}^{\mathrm{f}}. \label{eq:Rfbar}
\end{equation}  
from \eqref{eq:channel-UE-IRS}. Then,  $\boldsymbol{\mathcal{A}}_{k\ell,nn^{\prime}}$ is computed as
\begin{align}
&\boldsymbol{\mathcal{A}}_{k\ell,nn^{\prime}}=\left[\overline{\vect{R}}_{k\ell}^{\mathrm{f}}\right]_{nn^{\prime}}\Bigg(\left[\bar{\vect{G}}_{\ell,1}\right]_{:n}\left[\bar{\vect{G}}_{\ell,1}\right]_{:n^{\prime}}^{\Htran} \nonumber\\
&+\underbrace{\mathbb{E}\left\{\left[\tilde{\vect{G}}_{\ell}\right]_{:n}\left[\tilde{\vect{G}}_{\ell}\right]_{:n^{\prime}}^{\Htran}\right\}}_{\boldsymbol{\mathcal{B}}_{\ell,nn^{\prime}}}\Bigg) \nonumber\\
& +\!\left[\overline{\vect{R}}_{k\ell}^{\mathrm{f}}\right]_{nn^{\prime}}\!\!\underbrace{\mathbb{E}\left\{\!\! \left(\!\sum_{s=2}^{S_{\ell}^{\mathrm{G}}}e^{\imagunit\theta^{\mathrm{G}}_{\ell,s}}\!\left[\bar{\vect{G}}_{\ell,s}\right]_{:n}\!\!\right)\!\!\left(\!\sum_{s^{\prime}=2}^{S_{\ell}^{\mathrm{G}}}\!\left[\bar{\vect{G}}_{\ell,s^{\prime}}\right]_{:n^{\prime}}^{\Htran}\!e^{-\imagunit\theta^{\mathrm{G}}_{\ell,s^{\prime}}}\!\right)\!\!\right\}}_{\boldsymbol{\mathcal{C}}_{\ell,nn^{\prime}}} \label{eq:RbarGf}
\end{align} 
from \eqref{eq:channel-BS-IRS}.  The  expectation $\boldsymbol{\mathcal{B}}_{\ell,nn^{\prime}}$ can be evaluated by noting that 
\begin{equation}
\left[\tilde{\vect{G}}_{\ell}\right]_{:n} = \left(\vect{R}_{\ell}^{\mathrm{G,BS}}\right)^{\frac12}\vect{W}_{\ell}\left[\left(\vect{R}_{\ell}^{\mathrm{G,RIS}}\right)^{\frac12}\right]_{:n} \label{eq:Gtilde2}
\end{equation}
from  \eqref{eq:Gtilde}. Since the elements of $\vect{W}_{\ell}$ are i.i.d. standard complex Gaussian random variables,
 inserting \eqref{eq:Gtilde2} into the $\boldsymbol{\mathcal{B}}_{\ell,nn^{\prime}}$, we have
\begin{align}
\boldsymbol{\mathcal{B}}_{\ell,nn^{\prime}}=&\mathbb{E}\Bigg\{ \left(\vect{R}_{\ell}^{\mathrm{G,BS}}\right)^{\frac12}\vect{W}_{\ell}\left[\left(\vect{R}_{\ell}^{\mathrm{G,RIS}}\right)^{\frac12}\right]_{:n}\nonumber\\
&\times\left[\left(\vect{R}_{\ell}^{\mathrm{G,RIS}}\right)^{\frac12}\right]_{n^{\prime}:}\vect{W}_{\ell}^{\Htran}\left(\vect{R}_{\ell}^{\mathrm{G,BS}}\right)^{\frac12}\Bigg\} \nonumber \\
=&\sum_{m=1}^{N}\sum_{m^{\prime}=1}^{N}\mathbb{E}\Bigg\{ \left(\vect{R}_{\ell}^{\mathrm{G,BS}}\right)^{\frac12}\left[\left(\vect{R}_{\ell}^{\mathrm{G,RIS}}\right)^{\frac12}\right]_{mn}\nonumber\\
&\times\left[\left(\vect{R}_{\ell}^{\mathrm{G,RIS}}\right)^{\frac12}\right]_{n^{\prime}m^{\prime}}\left[\vect{W}_{\ell}\right]_{:m} \left[\vect{W}_{\ell}\right]_{:m^{\prime}}^{\Htran}\left(\vect{R}_{\ell}^{\mathrm{G,BS}}\right)^{\frac12}\Bigg\} \nonumber \\
&\hspace{-14mm}=\sum_{m=1}^{N}\!\! \left(\vect{R}_{\ell}^{\mathrm{G,BS}}\right)^{\frac12}\!\!\left[\!\left(\vect{R}_{\ell}^{\mathrm{G,RIS}}\right)^{\frac12}\!\right]_{\!mn}\!\!\left[\left(\vect{R}_{\ell}^{\mathrm{G,RIS}}\right)^{\frac12}\!\right]_{\!n^{\prime}m}\!\!\!\left(\vect{R}_{\ell}^{\mathrm{G,BS}}\right)^{\frac12}\nonumber\\
=&\left[\vect{R}_{\ell}^{\mathrm{G,RIS}}\right]_{n^{\prime}n}\vect{R}_{\ell}^{\mathrm{G,BS}} \label{eq:expectation} 
\end{align}
where $\left[\left(\vect{R}_{\ell}^{\mathrm{G,RIS}}\right)^{\frac12}\right]_{n^{\prime}:}$ denote the $n^{\prime}$th row of the matrix $\left(\vect{R}_{\ell}^{\mathrm{G,RIS}}\right)^{\frac12}$ and in the first equality we have used that this matrix is Hermitian symmetric. 

The second expectation in \eqref{eq:RbarGf}, which is $\boldsymbol{\mathcal{C}}_{\ell,nn^{\prime}}$ is given by
\begin{equation}
\boldsymbol{\mathcal{C}}_{\ell,nn^{\prime}} = \sum_{s=2}^{S_{\ell}^{\mathrm{G}}} \left[\bar{\vect{G}}_{\ell,s}\right]_{:n}\left[\bar{\vect{G}}_{\ell,s}\right]_{:n^{\prime}}^{\Htran}
\end{equation}
from the independence of the phase-shifts and the fact that $\mathbb{E}\{e^{j\theta}\}=0$ for a random variable $\theta \sim\mathcal{U}[0,2\pi)$. So, $\boldsymbol{\mathcal{A}}_{k\ell,nn^{\prime}}$ in \eqref{eq:RbarGf} is given by
\begin{align}
&\boldsymbol{\mathcal{A}}_{k\ell,nn^{\prime}}=\left[\overline{\vect{R}}_{k\ell}^{\mathrm{f}}\right]_{nn^{\prime}}\nonumber\\
&\times\Bigg(\sum_{s=1}^{S_{\ell}^{\mathrm{G}}} \left[\bar{\vect{G}}_{\ell,s}\right]_{:n}\left[\bar{\vect{G}}_{\ell,s}\right]_{:n^{\prime}}^{\Htran}+\left[\vect{R}_{\ell}^{\mathrm{G,RIS}}\right]_{n^{\prime}n}\vect{R}_{\ell}^{\mathrm{G,BS}} \Bigg).
\end{align}
Inserting $\boldsymbol{\mathcal{A}}_{k\ell,nn^{\prime}}$ into \eqref{eq:Rklr2}, we complete the proof.

\section{Proof of Lemma~\ref{lemma4}} \label{appendix2}

Assuming that the angular deviations are sufficiently small such that $\cos(\delta)\approx 1$, $\cos(\epsilon)\approx 1$, $\sin(\delta)\approx \delta$, $\sin(\epsilon)\approx \epsilon$, $\left[\vect{R}\right]_{ml}$ in \eqref{eq:ml-R} can be approximated as
\begin{align}
&\left[\vect{R}\right]_{ml} 
\stackrel{(a)}{\approx}\underbrace{e^{\imagunit2\pi d_{{\rm H}}^{ml}\sin(\varphi)\cos(\theta)}e^{\imagunit2\pi d_{{\rm V}}^{ml}\sin(\theta)}}_{\triangleq A_{ml}} \nonumber\\
&\times \int \underbrace{e^{\imagunit{2\pi d_{{\rm H}}^{ml}\cos(\varphi)\cos(\theta)}\delta}}_{\triangleq e^{\imagunit B_{ml}\delta}} \frac{1}{\sqrt{2\pi}\sigma_{\varphi}}e^{-\frac{\delta^2}{2\sigma_{\varphi}^2}}d\delta  \nonumber\\
&\times \int \underbrace{e^{\imagunit2\pi\Big( -d_{{\rm H}}^{ml}\cos(\varphi)\sin(\theta)\delta-d_{{\rm H}}^{ml}\sin(\varphi)\sin(\theta)+ d_{{\rm V}}^{ml}\cos(\theta)\Big)\epsilon}}_{\triangleq e^{\imagunit \left(C_{ml}\delta+D_{ml}\right)\epsilon}}\nonumber\\
&\times\frac{1}{\sqrt{2\pi}\sigma_{\theta}}e^{-\frac{\epsilon^2}{2\sigma_{\theta}^2}} d\epsilon \nonumber \\
\stackrel{(b)}{=}&\frac{ A_{ml} \widetilde{\sigma}_{ml}}{\sigma_{\varphi}} e^{\frac{D_{ml}^2\sigma_{\theta}^2\left(C_{ml}^2\sigma_{\theta}^2\widetilde{\sigma}_{ml}^2-1\right)}{2}} \nonumber \\
&\times \int e^{\imagunit B_{ml}\delta}\frac{1}{\sqrt{2\pi}\widetilde{\sigma}_{ml}}e^{-\frac{\left(\delta+C_{ml}D_{ml}\sigma_{\theta}^2\widetilde{\sigma}_{ml}^2\right)^2}{2\widetilde{\sigma}_{ml}^2}} d\delta \nonumber \\
\stackrel{(c)}{=}&\frac{ A_{ml} \widetilde{\sigma}_{ml}}{\sigma_{\varphi}} e^{\frac{D_{ml}^2\sigma_{\theta}^2\left(C_{ml}^2\sigma_{\theta}^2\widetilde{\sigma}_{ml}^2-1\right)}{2}}  e^{-\imagunit B_{ml}C_{ml}D_{ml}\sigma_{\theta}^2\widetilde{\sigma}_{ml}^2}e^{-\frac{B_{ml}^2\widetilde{\sigma}_{ml}^2}{2}} \label{eq:spatial_correlation_UPA2}
\end{align}
where we have used the trigonometric identities $\sin(A+B)=\sin(A)\cos(B)+\cos(A)\sin(B)$, $\cos(A+B)=\cos(A)\cos(B)-\sin(A)\sin(B)$, and the approximations $\cos(\delta)\approx 1$, $\cos(\epsilon)\approx 1$, $\sin(\delta)\approx \delta$, and $\sin(\epsilon)\approx \epsilon$ for sufficiently small angular deviations $|\delta|$ and $|\epsilon|$ in $(a)$. In $(b)$, we use the fact that the characteristics function for the real Gaussian variable $x\sim\mathcal{N}(\mu,\sigma^2)$ is given by $\mathbb{E}\left\{e^{\imagunit\omega x}\right\}=e^{\imagunit\omega\mu-\frac{\omega^2\sigma^2}{2}}$. Moreover, we arrange the equation to obtain a Gaussian probability density function after the term $ e^{\imagunit B_{ml}\delta}$ with the variance given in \eqref{eq:sigmatilde}. Using the same identity  $\mathbb{E}\left\{e^{\imagunit\omega x}\right\}=e^{\imagunit\omega\mu-\frac{\omega^2\sigma^2}{2}}$ for $x\sim\mathcal{N}(\mu,\sigma^2)$ in $(c)$, we obtain the end result.

\bibliographystyle{IEEEtran}
\bibliography{IEEEabrv,refs}

\end{document}